\documentclass[11pt]{article}
\usepackage{amsmath}
\usepackage{amsfonts}
\usepackage{graphicx}
\usepackage{cancel}
\usepackage{fullpage}
\usepackage{hyperref}
\usepackage{color}
\usepackage[format=plain,textfont=it]{caption}

\newcommand{\deltabar}{\delta\hspace*{-0.2em}\bar{}\hspace*{0.1em}}
\newcommand{\Lagr}{\mathcal{L}}
\newcommand{\order}[1]{\mathcal{O}\!\left(#1\right)}

\newcommand{\derOrd}[2]{\frac{d#1}{d#2}}
\newcommand{\der}[2]{\frac{\partial#1}{\partial#2}}

\newcommand{\non}{\nonumber\\}
\title{Free energy from forward scattering in 1+1d}
\date{ \textit{\small Initiative for the Theoretical Sciences\\  The Graduate Center, CUNY\\ 365 Fifth Ave, New York, NY 10016, USA}
}
\author{Daniel Schubring}
\numberwithin{equation}{section}

\begin{document}
\maketitle

	\noindent The free energy, or equivalently the ground state energy in finite volume, may be calculated from forward scattering amplitudes using a formula due to Dashen, Ma, and Bernstein \cite{dmb}. However a naive treatment leads to singularities when considering the scattering of three or more particles. It is shown in detail how the approach can be applied to multi-particle scattering in various massive scalar theories in 1+1d, with or without integrability. The results for the sinh-Gordon, Lieb-Liniger, and $O(N)$ non-linear sigma models are compared to exact results. It is shown how bound states can be considered in this approach by considering the attractive Lieb-Liniger model.
	
	\tableofcontents
	\section{Introduction}
			One of the most basic physical quantities that may be calculated from a quantum field theory is the free energy as a function of temperature. It is well-understood how to calculate this perturbatively using compactified Euclidean time, an approach referred to here as thermal field theory. The free energy should in principle encode much information about the zero-temperature dynamics of the theory, but since the size $\beta$ of the thermal circle is involved in each propagator of a diagram, the thermal field theory approach mixes statistics with dynamics in an essential way.

Much better in this respect is the thermodynamic Bethe ansatz (TBA) \cite{yangYangTBA}. The zero-temperature dynamics are encoded in kernels $K$ which involve derivatives of the logarithm of the two-body S-matrix (see e.g. equation \eqref{Def K LL}). These kernels then appear in an integral equation to determine the free energy. However, this approach only works for integrable systems.
	
	There is a less widely known third approach to calculate the free energy introduced by Dashen, Ma, and Bernstein (DMB) \cite{dmb} which extends a formula by Beth and Uhlenbeck \cite{bethUhlenbeck}. This approach explicitly separates the zero temperature dynamics from the temperature by writing the partition function in terms of the density of states $\rho(E)$,
	$$Z=\int dE e^{-\beta E}\rho(E).$$
	The density of states may be calculated in terms of forward scattering amplitudes at zero-temperature, even for non-integrable systems. In fact the expression for $\rho$ involves a derivative of a logarithm of an S-matrix operator, and it is suggestive of the kernel $K$ in the TBA approach,
	\begin{align}
		\rho(E)=\rho^{(0)}(E)+\frac{1}{2\pi i}\derOrd{}{E}\text{Tr}\log \hat{S}(E). \label{Eq DMB formula}
	\end{align}
	Here $\rho^{(0)}$ refers to the non-interacting density of states, and the operator $\hat{S}$ will be defined carefully in Sec \ref{Sec Intro DMB} where the derivation of the DMB formula is reviewed.
	
	The calculation of $\rho$ in the DMB approach has certain subtleties, and the lack of explicit examples in the literature has perhaps hindered its adoption.  In this paper, in order to clearly illustrate the DMB approach we will consider certain massive scalar field theories in two spacetime dimensions for which we can compare with calculations in the thermal field theory and TBA approaches.

\subsection{Extending the L\"{u}scher formula via Beth-Uhlenbeck}
A reconsideration of the DMB approach \cite{dmb, dm1970, dm1971, norton1986elementary}, as well as related approaches to calculating the self-energy and higher correlation functions \cite{luscher1984,luscher1986 I,jeon1993, jeonEllis1998}, may ultimately be useful for extracting information on multi-particle scattering  amplitudes from lattice simulations. Note that in 1+1d the free energy may alternatively be understood as a calculation of the ground state energy as a function of finite volume $L$. A better understanding of how to relate non-integrable scattering data to the energy levels as a function of $L$ is relevant for the effective theory of Yang-Mills flux tubes \cite{athenodorouEtAl 3plus1, athenodorouEtAl 2plus1, aharonyKomargodski2013}. There has been great success in using the TBA to relate the two-body S-matrix on the worldsheet of the flux tube to lattice calculations of the energy levels of the flux tube as a function of $L$ \cite{dubovskyFlaugerGorbenko2013,dubovskyFlaugerGorbenko2015}. Even so, it is understood that the worldsheet theory for both 3D and 4D Yang-Mills is not actually integrable \cite{cooperEtAlConfiningStringIntegrability,chenEtAlFluxTubesTTbar2018}.
	
	There is an alternate approach due to L\"{u}scher \cite{luscher1986 II,luscherWolff1990} which is well-known in this context, and which can be applied to the two-body energy levels in non-integrable systems. As will be explained below, the present investigation is an extension of the infinite volume limit of the L\"{u}scher formula to multi-particle scattering states (see also \cite{hansenSharpe2014}). Although the effective string theory program is a motivating example, in this work we will focus on models with massive particles in order to better understand how multi-particle scattering affects the finite size dependence of the energy levels in a simple case. However note that shortly after publication of the initial preprint of this article on arXiv, an independently developed work \cite{BaratellaEliasMiroGendy2024} appeared that directly treated the integrable worldsheet theory for $D=3$, as well as the lowest order correction to the QCD thermal partition function.

	The L\"{u}scher formula in two dimensions is essentially the same thing as the two-particle asymptotic Bethe ansatz (ABA)
	\begin{align}
		2\pi i n = i k_n L + 2\log S(k_n),
	\end{align}
where $n$ is an integer specifying the energy level\footnote{Since the equation is written only in terms of the relative momentum, $n$ varies in steps of $2$ for fixed center of mass momentum.}, $k_n$ is the relative momenta between the particles, and $S(k_n)$ is the two-body S-matrix. Strictly speaking this formula is valid for massive particles in the limit of large $mL$, and any additional terms which vanish in this limit are suppressed in this section. Since the energy levels are closely spaced in this limit, it makes sense to instead consider the density of two-particle states $\rho_{(2)}$,
\begin{align*}
	\rho_{(2)}(E)= \left(\frac{dE}{dk}\right)^{-1}\frac{1}{k_{n+2}-k_n}= \frac{L}{4\pi} \left(\frac{dE}{dk}\right)^{-1} +\frac{1}{2\pi i}\frac{d}{dE}\log S(E).
\end{align*}
Here we are taking fixed center of mass momentum, so that the total energy $E$ is a single-variable function of the relative momentum $k$, and the two body S-matrix is rewritten in terms of $E$. The first term is just the density of two-particle states in the non-interacting theory, and the remaining term $\delta\rho_{(2)}$ is
		\begin{align}
		\delta\rho_{(2)}(E)  = \frac{1}{2\pi i}\derOrd{}{E}\log S(E).\label{beth uhlenbeck}
	\end{align}
This formula was originally derived in ordinary quantum mechanics by Beth and Uhlenbeck \cite{bethUhlenbeck}.

Of course, the density of states also has contributions from states with more than two particles. For integrable systems we may continue to use the ABA in order to calculate the density of states \cite{katoWadati2001}. But such an approach is not directly applicable to non-integrable systems.

Instead we can extend the Beth-Uhlenbeck formula following DMB \cite{dmb}. The DMB formula \eqref{Eq DMB formula} is superficially identical to the Beth-Uhlenbeck formula, but as will be discussed further in Sec \ref{Sec Intro DMB} the operator $\hat{S}(E)$ is an S-matrix operator that may be applied to arbitrary multiparticle states, and the $E$ argument need not be equal to the energy of the states involved in the trace.
		
	\subsection{Forward scattering singularities}
	
		To see roughly how \eqref{Eq DMB formula} is to be interpreted, write the non-trivial part of the S-matrix operator as usual in terms of a T-matrix operator with an energy delta function factored out,
	\begin{align}
		\hat{S}(E)=1-2\pi i \delta(E-H_0)\hat{T}(E).\label{def S matrix operator}
	\end{align}
	The factor $\log \hat{S}(E)$ may be expanded to first order in powers of $\hat{T}$, and after integrating by parts in $E$ the partition function may be written as
	\begin{align*}
		\delta Z &= -\beta\int d\alpha \,e^{-\beta E_\alpha}T_{\alpha\alpha}+\order{T^2}.
	\end{align*}
	Here the trace was taken over eigenstates $\alpha$ of $H_0$ with energy $E_\alpha$, and $d\alpha$ schematically represents the appropriate measure for the trace. $T_{\alpha\alpha}$ is the forward scattering amplitude with both an incoming and outgoing state of $\alpha$.	As usual, we may consider instead the connected part of the T-matrix, $T^c$, and the free energy density\footnote{We use $L$ for spatial volume, although the following formulas apply to arbitrary dimension.} $f=-(\beta L)^{-1}\log Z$,
	\begin{align}
		\delta f= L^{-1}\int d\alpha \,e^{-\beta E_\alpha}T^c_{\alpha\alpha}+\order{T^2}.\label{Eq DMB formula single T}
	\end{align}
	
	Here already we see a potential issue. A connected T-matrix amplitude $T^c_{\beta \alpha}$ may involve various delta functions, and in the forward scattering limit $\beta=\alpha$ the argument of these delta functions may go to zero. One delta function will just be due to momentum conservation and it will lead to an overall factor of $L$ to cancel the $L^{-1}$ in \eqref{Eq DMB formula single T}. But additional delta functions may arise from internal propagators in scattering amplitudes which can go on-shell. 
	
\begin{figure}
		\centering
		\includegraphics[width=0.25\textwidth]{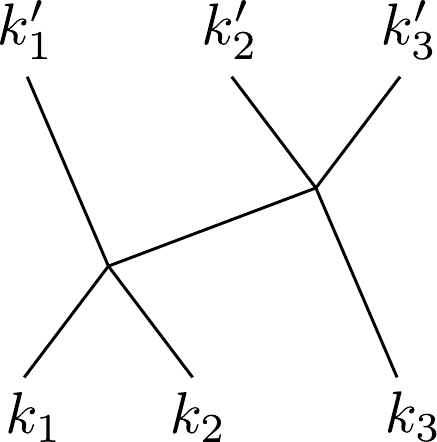}
		\caption{A three-body scattering amplitude which may go on-shell in the forward scattering case.}\label{Fig schematic}
	\end{figure}

	For instance the lowest-order three-particle scattering amplitude in a relativistic scalar field theory in Fig \ref{Fig schematic} is just proportional to the internal propagator
	$$\frac{1}{(E_1+E_2-E_1')^2-(k_1+k_2-k_1')^2-m^2+i\epsilon},$$
	where the $E$ are on-shell energies $E=\sqrt{k^2+m^2}$ associated to the spatial momentum $k$ of a given external leg. This propagator goes on-shell when $k_1'$ equals $k_1$ or $k_2$. The propagator can be broken into a real principal part, and an imaginary delta function. These delta functions can not be ignored and in fact for integrable models they are necessary to ensure that the set of final momenta are equal to the initial ones (see the discussion in \cite{dorey1996}). So the forward scattering amplitude $T^c_{\alpha \alpha}$ will certainly involve delta functions with a vanishing argument, and these forward scattering singularities need to be interpreted for the DMB formula to make sense.
	
	As we will show perturbatively in Sec \ref{Sec Intro DMB mult part}, this issue is resolved by including the higher order powers of $T$ in the expansion in \eqref{Eq DMB formula single T}. The delta function parts of the internal propagators in diagrams like Fig \ref{Fig schematic} are canceled, and only the principal part remains. This introduces some new questions since for three-particle scattering in integrable theories the principal part is taken to vanish. Nevertheless we will show that there are two subtle types of contributions to the free energy due to the principal part: \emph{on-shell contributions} and  \emph{collinear contributions}.
	

	On-shell contributions are due to forward scattering diagrams that would have had an on-shell divergence, such as Fig \ref{Fig schematic} with $k_1'$ taken to be $k_1$ or $k_2$. The principal part of the internal propagator vanishes exactly on-shell, but it is large slightly off-shell, and the slightly off-shell region will contribute since the energy delta function involved in the S-matrix operator \eqref{def S matrix operator} actually has a width that depends on $i\epsilon$.
	
	The calculation of the on-shell contributions has previously been discussed to a certain extent for the scattering of three or four particles \cite{dm1970,dm1971}, as well as more generally using a different approach \cite{norton1986elementary}. In section \ref{Sec On shell} we give an arguably simpler calculation of the on-shell contributions for an arbitrary number of particles in relativistic theories from both the thermal field theory and DMB perspectives.
	
	Besides the on-shell contributions, there are also collinear contributions which appear at lowest order due to forward scattering diagrams such as Fig \ref{Fig schematic} with $k_1' = k_3$, so that the internal propagator need not be on-shell. If $k_1, k_2, k_3$ are all distinct then the principal part vanishes as per the usual argument. But since we are integrating over all momentum in the DMB formula, we need to consider the cases where the particle momenta become collinear. It turns out that the principal part doesn't vanish in a small $i\epsilon$ dependent region around the collinear point $k_1=k_2=k_3$, and this is actually a significant contribution to the free energy.
	
	 In the $\phi^4$ theory there are two free energy diagrams at second order in the thermal field theory approach,  which may be called the \emph{bubble} and \emph{melon} diagrams (see Fig \ref{Fig bubble melon}). We will show that the melon diagram in 1+1d is entirely the result of collinear contributions. Note that the massive \cite{bugrijShadura1995,massiveBasketball} and massless \cite{frenkelSaaTaylor1992} $\phi^4$ theory have been investigated before in 3+1d via the DMB approach and the related approach of Barton \cite{barton1990finite}, but the essential features of the 1+1d case were not seen.  The 1+1d melon diagram is calculated in Sec \ref{Sec Intro diagrams bubble melon} and compared to a perturbative expansion of the TBA in Sec \ref{Sec TBA perturbative}.  In Sec \ref{Sec TBA LL} it is shown how these collinear contributions are necessary in perturbation theory since there are bound state contributions to the free energy upon flipping the sign of the coupling constant.
	
\subsection{Outline} A brief outline of this paper is as follows. In Sec \ref{Sec Intro DMB} we will review the general derivation of the DMB formula and apply it to integrable models involving a single massive scalar such as sinh-Gordon. In Sec \ref{Sec Intro diagrams} we consider sinh-Gordon and the non-integrable $\phi^4$ theory at second order in perturbation theory, and show how free energy diagrams in the thermal field theory approach are connected to two- or three-body forward scattering diagrams in the DMB approach.
		
		In Sec \ref{Sec On shell} diagrams involving only on-shell contributions are calculated at arbitrary order in perturbation theory, and it is shown how these give the leading contribution to the free energy for the large $N$ extension of the $\phi^4$ model. In Sec \ref{Sec TBA} we investigate the treatment of bound states in the DMB approach by considering the non-relativistic Lieb-Liniger model with attractive coupling. We show how the collinear contributions at repulsive coupling are related to the contributions due to bound states for attractive coupling. In Sec \ref{Sec Discussion} we discuss future investigations into the DMB approach.
		
\section{The DMB formula and integrable models}\label{Sec Intro DMB}

\subsection{Derivation of the DMB formula}
First let us review some material from scattering theory. The Lippmann-Schwinger equation expresses the in and out eigenstates $\Psi^{\pm}$ of the Hamiltonian $H=H_0+V$ in terms of the eigenstates $\Phi$ of $H_0$ with the same eigenvalue $E$,
\begin{align}
	\Psi^\pm = \Phi+\frac{1}{E-H_0\pm i\epsilon}V\Psi^{\pm}.
\end{align}

This can be written in terms of a quantity $G_0$ that is usually referred to as the propagator in this context,
\begin{align}
	G_0(E)\equiv \frac{1}{E-H_0+ i\epsilon}.
\end{align}
An interacting propagator $G$ may also be defined,
\begin{align}
	G(E)\equiv \frac{1}{E-H+ i\epsilon}=\frac{1}{1-G_0(E)V}G_0(E).
\end{align}

Now we can define a $T$-matrix operator so that the Lippmann-Schwinger equation can be written as $V\Psi^+=\hat{T}(E)\Phi $,
\begin{align}
	\hat{T}(E)\equiv V\frac{1}{1-G_0(E) V}.\label{Def T}
\end{align}
An $S$-matrix operator may also be defined in terms of $T$,
\begin{align}
	\hat{S}(E)&\equiv 1+\left(G_0(E)-G_0(E)^\dagger\right)\hat{T}(E)\label{Def S}\\
	&=\left(1-G_0(E)^\dagger V\right) \left(1-G_0(E) V\right)^{-1}
\end{align}
The first line is just the usual relation between $S$ and $T$ since $G_0-G_0^\dagger \rightarrow -2\pi i \delta(E-H_0)$ in the $\epsilon\rightarrow 0$ limit.

The $\hat{S}(E)$ operator is defined even outside the usual limits in which scattering theory applies, but in order to interpret it as the usual $S$-matrix it must be acting on an $H_0$ eigenstate $\Phi$ with the same eigenvalue $E$ as its argument, and the infinite volume and  $\epsilon\rightarrow 0$ limits must be taken. Keeping these caveats in mind, we will suppress the $E$ argument except when necessary.

Now the DMB formula follows from some simple manipulations using the formulas above and the fact that $\der{}{E}G_0(E)= -G_0(E)^2$.
\begin{align}
	\frac{1}{2\pi i}\derOrd{}{E}\text{Tr}\log \hat{S} &= 		\frac{1}{2\pi i}\derOrd{}{E}\text{Tr}\left[\log\left(1-G_0^\dagger V\right)-\log\left(1-G_0 V\right)\right]\non
	&=	\frac{1}{2\pi i}\text{Tr}\left[\frac{\left(G_0^\dagger\right)^2V}{1-G_0^\dagger V}-\frac{G_0^2V}{1-G_0 V}\right]\non
	&=	\frac{1}{2\pi i}\text{Tr}\left[\left(\frac{1}{1-G_0^\dagger V}-1\right)G_0^\dagger-\left(\frac{1}{1-G_0 V}-1\right)G_0\right]\non
	&=-	\frac{1}{2\pi i}\text{Tr}\left[\left(G-G^\dagger\right)-\left(G_0-G_0^\dagger\right)\right]\non
	&= \text{Tr}\left[\delta(E-H)-\delta(E-H_0)\right],
\end{align}
where the last line holds in the $\epsilon\rightarrow 0$ limit. This is now in the form of the difference between the density of states $\rho(E)-\rho^{(0)}(E)$ as required in the DMB formula \eqref{Eq DMB formula}.

In situations where there is a conserved particle number $N$ in both the free and interacting theories it further makes sense to introduce a chemical potential $\mu$ and corresponding \emph{fugacity} $y\equiv e^{\beta \mu}$, in terms of which the grand canonical partition function is
\begin{align}
	\delta Z=\sum_{N=2}^\infty y^N \int dE e^{-\beta E}\frac{1}{2\pi i}\derOrd{}{E}\text{Tr}_{(N)}\log \hat{S}.\label{Eq DMB formula grand canonical}
\end{align}

In what follows fugacity dependence will be used as a counting device to single out the trace over $N$-particle states $\text{Tr}_{(N)}$, and we will set $y=1$ at the end. In the theories of massive particles considered here the term in the partition function proportional to $y^N$ will have an asymptotic temperature dependence on the order of $\sim e^{-N \beta m}$. So the fugacity expansion may equivalently be understood as a low temperature expansion in powers of $e^{-\beta m}$.

We will use a notation where a subscript in parenthesis denotes the order in the fugacity expansion. On the other hand, a superscript in parentheses denotes the order in a coupling constant ($c$ or $g^2$) expansion, and a $\delta$ prefix denotes the difference of a quantity from the free theory. Later on we will introduce an $n$-body expansion that takes into account particle statistics, and this will be indexed by a subscript without parentheses.
\subsection{Integrable models: Sinh-Gordon and Lieb-Liniger}
It should be clear that the DMB formula above is quite general since it simply follows from the definitions of the operators $\hat{T}$ and $\hat{S}$, but to illustrate it in practice we consider the sinh-Gordon model (SG) \cite{sinhGordonOrig}, with Lagrangian
\begin{align}
	\Lagr_{SG} = \frac{1}{2}\left(\partial \phi\right)^2+\frac{\left(m^2+\delta m^2\right)}{g^2}\left(\cosh\left( g\phi\right) - 1\right).\label{Lagr sinh Gordon}
\end{align}
Here $\delta m^2$ is a counterterm set to ensure that $m$ is the physical mass.

In many respects the non-relativistic limit of SG may be understood to be the Lieb-Liniger model (LL) \cite{LiebLiniger1963}, which involves bosons in ordinary quantum mechanics interacting through a repulsive two-body delta function potential $$V_{LL}(x)=\frac{c}{m}\delta(x).$$
The parameter $g$ in the SG Lagrangian is related to the parameter $c$ by
\begin{align}
	c=m\sin \frac{\pi g^2}{8\pi+g^2}  = \frac{m g^2}{8}-\frac{m g^4}{64\pi}+\order{g^6}.
\end{align}

Both SG and LL are integrable and the exact two-body S-matrix is given by
\begin{align}
	S(p)=\frac{p-i c}{p+ic}.\label{exact two body S matrix}
\end{align}
In LL, $p$ is taken to be the momentum difference between the particles, whereas in SG, $p = m \sinh \theta$ with $\theta$ being the rapidity difference. 

Since both models involve a single particle type with no internal indices and no bound states, the TBA takes the simplest possible form (see \cite{tbaIntroTongeren} for a review). The free energy density $f$ is calculated as
\begin{align}
	f= -T\int \frac{dp}{2\pi} \log\left(1+e^{-\epsilon(p)/T}\right)\label{TBA free energy}
\end{align}
where the pseudo-energy $\epsilon$ is determined by solving the TBA equation
\begin{align}
	\epsilon=E-\mu-T\log \left(1+e^{-\epsilon/T}\right)\star K.\label{TBA equation}
\end{align}
Here $E$ is the single particle dispersion relation, and the $\star$ denotes convolution\footnote{The convolution is taken over the rapidity variable and the momentum variable in SG and LL, respectively.}
$$f\star g (\theta)\equiv \int d\theta' f(\theta')g(\theta-\theta').$$
$K$ is a kernel that is derived from the exact S-matrix. The kernels for LL and SG are respectively,
\begin{gather}
	K_{LL} = \frac{1}{2\pi i}\frac{d}{dp}\log S(p)=\frac{1}{\pi}\frac{c}{p^2+c^2},\label{Def K LL}\\
	K_{SG} = \frac{1}{2\pi i}\frac{d}{d\theta}\log S\left(m\sinh\theta\right)=m\cosh\theta K_{LL}.\label{Def K SG}
\end{gather}
The formal similarity of the kernels to the DMB formula for the density of states is not an accident, since both approaches will reduce to the Beth-Uhlenbeck formula \eqref{beth uhlenbeck} in the two-particle limit, as shown in the section below.
\subsection{Applying DMB to integrable models}\label{Sec DMB diagonal S matrix}
\label{Section diagonal S matrix}

Since SG and LL are integrable and only involve one species of particle, this implies that arbitrary multiparticle states $\alpha$ in the momentum basis are eigenstates of the S-matrix operator
$$\hat{S}(E_\alpha)|\alpha\rangle = S(\alpha)|\alpha\rangle.$$
Here $\hat{S}$ refers to the operator appearing in the DMB formula, and $S(\alpha)$ is a complex eigenvalue which ends up being factorizable into two-body S-matrices.

The normalization of the eigenstates is fixed to be compatible with the measure in the sense that $\langle \alpha|\beta\rangle$ acts like a delta function,
$$\int d\alpha \langle \alpha|\beta\rangle f(\alpha) =f(\beta).$$

Now let us use these eigenstates as the basis for the trace in the DMB formula,
\begin{align}
	\delta Z&=\frac{\beta}{2\pi i}\int dE e^{-\beta E}\text{Tr}\log \hat{S}(E)\non&=-\frac{\beta}{2\pi i}\sum_{k=1}\frac{1}{k}\int dE e^{-\beta E}\int d\alpha \langle \alpha|\left(2\pi i \delta(E-H_0)\hat{T}(E)\right)^k|\alpha\rangle\non
	&=-{\beta}\sum_{k=1}\frac{1}{k}\int d\alpha\, e^{-\beta E_\alpha}\langle \alpha|\hat{T}(E_\alpha)\left(2\pi i \delta(E_\alpha-H_0)\hat{T}(E_\alpha)\right)^{k-1}|\alpha\rangle.
\end{align}
Here the last step of using one delta function to evaluate the $E$ integral is somewhat questionable since the delta function is really $G_0-G_0^\dagger$ for finite $i\epsilon$, and there may be problems with the vanishing $i\epsilon$ limit in the T-matrix amplitudes. But for now let us proceed. This may be evaluated further by inserting an integration over a complete set of states between each $\delta(E-H_0)T$, and using the idea that
\begin{align}
	\langle \beta | 2\pi i \delta(E_\alpha-E_\beta)\hat{T}(E_\alpha) |\alpha\rangle = \left(1-S(\alpha)\right)\langle \beta |\alpha \rangle.\label{exact T matrix}
\end{align}
The result is
\begin{align}
	\delta Z		&=-\beta\sum_{k=1}\frac{1}{k}\int d\alpha d\beta_1\dots d\beta_{k-1} \,e^{-\beta E_\alpha}\langle \alpha|\hat{T}(E_\alpha)|\beta_1\rangle\dots\langle \beta_{k-1} | 2\pi i \delta(E_\alpha-E_{k-1})\hat{T}(E_\alpha) |\alpha\rangle \non&=\beta\int d{\alpha}\, e^{-\beta E_\alpha}\left(\frac{T(\alpha)}{1-S(\alpha)}\right)\log S(\alpha),\label{exact DMB attempt}
\end{align}
where $T(\alpha)\equiv \langle \alpha | \hat{T}(E_\alpha)|\alpha\rangle$ is the on-shell forward scattering T-matrix. Note that considering \eqref{exact T matrix}, $T(\alpha)$ appears to be proportional to $1-S(\alpha)$ and the factor in parenthesis in \eqref{exact DMB attempt} appears to just be a kinematic factor.

This formula is able to reproduce the correct free energy for two-particle states. A quick way to see this is to use a normalization of two-particle states such that $d\alpha = dE dP$ with $E$ and $P$ being the total energy and momentum. If it is valid to cancel the energy delta function from both sides of \eqref{exact T matrix} then we can calculate $T$,
\begin{align*}
	2\pi i T(\alpha)=\left(1-S(\alpha)\right)\frac{L}{2\pi}.
\end{align*} 
Then after integrating by parts, and rewriting in terms of the two-particle free energy density $f_{(2)}$, \eqref{exact DMB attempt} becomes

\begin{align}\delta f_{(2)}=-\frac{T}{2\pi}\int dP   dE  e^{-\beta E} \frac{1}{2\pi i} \der{}{E}\left.\log S(\alpha)\right|_{P}.\label{Eq f2 general}
\end{align}
This is just the Beth-Uhlenbeck formula \eqref{beth uhlenbeck} with the center of mass motion taken into account.

To write it more explicitly in SG, transform the integration measure to rapidity variables and use the kinematic relation $E^2=P^2+4m^2\cosh^2\theta$ to rewrite the $E$ derivative as a $\theta$ derivative,
\begin{align}
\delta f_{(2)}=-\frac{T}{2\pi}\frac{1}{2!}\int d\theta_1 d\theta_2  E  e^{-\beta E} \frac{1}{2\pi i} \der{}{\theta}\log S(\theta)= -T\int \frac{dp_1}{2\pi}e^{-\beta E_1}\int d\theta_2 e^{-\beta E_2}K_{SG}(\theta).\label{Eq f2 SG}
\end{align}
In the last equality this has been written in terms of a convolution integral over the SG kernel \eqref{Def K SG}, and it is a short step to see agreement with the TBA.

According to the formula for the grand canonical partition function \eqref{Eq DMB formula grand canonical}, the above expression should match the $\order{y^2}$ term in a fugacity expansion of free energy density \eqref{TBA equation}. This expansion may be carried out by expanding the pseudo-energy $\epsilon$ in powers of $y$ and solving the TBA equation \eqref{TBA equation} at each order. The result is
\begin{align}
	\delta f_{(2)} = \left(f-f^{(0)}\right)_{(2)}&= -T\int \frac{dp}{2\pi} e^{-\beta E}\left(e^{-\beta E}\star \delta K\right).\label{fugacity expansion y2}
\end{align}
Here we have subtracted off the the free energy density of a free boson gas,
\begin{align}f^{(0)}=+T\int \frac{dp}{2\pi} \log\left(1- e^{-\beta E}\right),\label{free boson gas}\end{align}
and $\delta K$ is the interacting part of the kernel,
\begin{align}
	\delta K_{LL}(p)\equiv K_{LL}(p)-\delta(p), \qquad \delta K_{SG}= m\cosh\theta\, \delta K_{LL}.
\end{align}

Clearly the TBA result \eqref{fugacity expansion y2} matches the corresponding DMB result for SG \eqref{Eq f2 SG}, and a similar agreement may be seen for the LL model. Agreement of the DMB formula with the TBA at $\order{y^2}$ has also been noted for the $O(N)$ model \cite{balogHegedusVirial2001}.

\subsection{Extending the argument to multiple particles}\label{Sec Intro DMB mult part}
Now we would like to apply the seemingly general formula \eqref{exact DMB attempt} to the scattering of three or more particles, but it is difficult to see how this could succeed. If $T/(1-S)$ is still just a kinematic factor, then since $S$ is factorizable into two-body S-matrices the $\log S$ factor just breaks up into a disconnected sum. In fact we were too cavalier in the treatment of the $i\epsilon$ terms in the derivation of \eqref{exact DMB attempt}, and the contribution due to multi-particle scattering depends sensitively on this.

The contribution to the free energy due to three-particle scattering may be investigated by expanding the TBA to the third order in fugacity,
\begin{align}
	\delta f_{(3)}	&= -T\int \frac{dp}{2\pi}e^{-\beta E} \left[e^{-\beta E}\left(e^{-\beta E}\star \delta K\right)+\frac{1}{2}\left(e^{-2\beta E}\star \delta K\right)\right]\non
	&\qquad -T\int \frac{dp}{2\pi}  e^{-\beta E}\left[\left(e^{-\beta E}\left(e^{-\beta E}\star \delta K\right)\right)\star \delta K+\frac{1}{2}\left(e^{-\beta E}\star \delta K\right)^2\right].\label{fugacity expansion y3}
\end{align}
This expression has also been derived by Kato and Wadati using an approach based on the asymptotic Bethe ansatz (ABA) \cite{katoWadati2001}.

As will be discussed more at the end of Sec \ref{Sec Intro diagrams single vertex}, the first line of \eqref{fugacity expansion y3} involving a single appearance of $\delta K$ is better understood as a correction to the two-body contribution \eqref{fugacity expansion y2} due to particle exchange. The two-body contribution $\delta f_2$ to all orders in fugacity is
\begin{align}
	\delta f_{2}=+T\int \frac{dp}{2\pi} \frac{1}{e^{\beta E}-1}\left[\log\left(1-e^{-\beta E}\right)\star \delta K\right].\label{TBA free energy two body}
\end{align}

The second line of \eqref{fugacity expansion y3} involving two factors of $\delta K$ represents genuine three-particle scattering. Contributions due to connected scattering of even more particles may be derived by continuing the fugacity expansion of the TBA, or using the ABA approach of \cite{katoWadati2001}. These expressions become increasingly complicated, and there is seemingly no way to derive them from the naive formula \eqref{exact DMB attempt}.

To better understand how multi-particle scattering is treated in the DMB formula, we consider the second order in perturbation theory, which is the lowest order for which three-particle scattering is relevant. The operator $\log \hat{S}$ may be expanded in powers of $V$,
\begin{align}
	\log \left(1+\left(G_0-\bar{G}_0\right)\hat{T}\right)^{(2)}&=\left(G_0-\bar{G}_0\right)V\frac{G_0+\bar{G}_0}{2}V\non
	&=\left(G_0-\bar{G}_0\right)\mathcal{P} \hat{T}^{(2)}.\label{DMB principal value}
\end{align}
Note that the internal propagator of $\hat{T}^{(2)}$ is 
$$G_0 = \frac{1}{E-H_0+i\epsilon} = \mathcal{P} G_0 - i\pi \delta(E-H_0).$$
For integrable theories, as long as all momenta of the particles are distinct (as is assumed in the ABA), the net contribution from the $\mathcal{P}G_0$ terms vanishes when summed over all diagrams. Rather it is the delta function term which leads to an $\hat{S}$ operator which is diagonal in the momentum basis, which is assumed in the derivation of \eqref{exact DMB attempt}. However \eqref{DMB principal value} asserts that only $\mathcal{P}G_0$ contributes in the DMB formula, and not the delta function part.

One way in which $\mathcal{P}T^{(2)}$ can be non-zero is if $E$ is not strictly set to be equal to the energy of the external particles in the scattering amplitude. This is possible since the ``delta function'' $(G_0-\hat{G}_0)$ has some width in terms of $\epsilon$. These type of contributions will be referred to as \emph{on-shell} contributions, since they involve diagrams for which the denominator of $G_0$ vanishes. On-shell contributions will be treated in DMB approach to arbitrary order in perturbation theory in Sec \ref{Sec On shell}.

The other way in which $\mathcal{P}T^{(2)}$ can be non-zero is if the particle momenta are collinear. This is not taken to be the case in the ABA, but it is encountered in the trace over free particle states in the DMB formula. We will treat the collinear contributions at second order in perturbation theory in Sec \ref{Sec Intro diagrams bubble melon} and the associated Appendix \ref{Sec Appendix melon}. As discussed in Sec \ref{Sec TBA LL}, the collinear contributions are closely connected to bound state contributions, and they are expected to be important at higher order in perturbation theory as well.

	\section{Cutting thermal field theory diagrams}\label{Sec Intro diagrams}
The DMB formula gives contributions to the free energy in terms of forward scattering amplitudes. In this section we will show how the forward scattering diagrams arise by cutting lines of vacuum bubble diagrams in thermal field theory. As a concrete example we will consider the second-order diagrams in sinh-Gordon \eqref{Lagr sinh Gordon} and its non-integrable truncation, the $\phi^4$ theory,
		\begin{align}
		\Lagr = \frac{1}{2}\left(\partial \phi\right)^2+\frac{m^2+\delta m^2 }{2}\phi^2+\frac{m^2g^2}{4!}\phi^4.\label{Lagr phi4}
	\end{align}
	\subsection{Single vertex corrections}\label{Sec Intro diagrams single vertex}

	\begin{figure}
		\centering
		\includegraphics[width=0.3\textwidth]{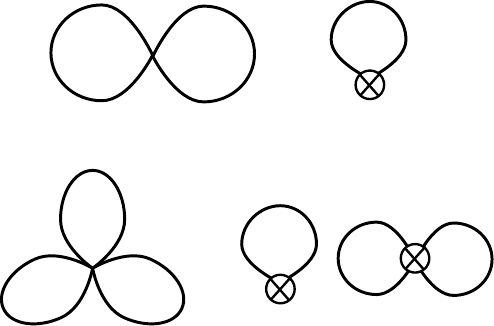}
		\caption{The free energy correction due to a $\phi^4$ vertex and an associated two-point counterterm (top), and a $\phi^6$ vertex and its associated counterterms (bottom).}\label{Fig first order}
	\end{figure}
	
	The simplest free energy corrections to consider involve a single vertex proportional to $\phi^{2n}$, which is contracted into a vacuum bubble diagram involving $n$ tadpoles. See Fig \ref{Fig first order} for the $n=2$ and $n=3$ cases, which are relevant in sinh-Gordon up to second-order. Besides the diagram involving the $\phi^{2n}$ vertex, there will also be diagrams at the same order involving counterterm vertices proportional to $\delta m^2$. For instance the total free energy density $f^{(1)}$ at first order is
	\begin{align}
		f^{(1)}=\frac{m^2g^2}{4!}3\left(I^2-I_0^2\right)+\frac{\delta m^2}{2}(I-I_0).\label{eq free energy 1st order}
	\end{align} 
	This is written in terms of the tadpole integral $I$,
	\begin{gather}
		I \equiv \int_\beta \frac{d^2 p}{\left(2\pi\right)^2}\frac{1}{p^2+m^2}=\frac{1}{\beta}\sum_n\int \frac{dk}{2\pi}\frac{1}{\omega_n^2+k^2+m^2},
	\end{gather}
	where the $\beta$ subscript shorthand indicates that Euclidean time is compactified, and the integral may be written explicitly in terms of Matsubara frequencies $\omega_n =2\pi n/\beta$. This integral may be written as a sum of a zero-temperature part $I_0$ and a thermal part that involves the bosonic occupation number $n_k$, which is defined as
	\begin{align}
		n_k(\beta)\equiv \frac{1}{e^{\beta E_k}-1}, \qquad 	E_k \equiv \sqrt{k^2+m^2}.\label{Def occupation number}
	\end{align}
	To make this decomposition, the sum over $\omega_n$ is transformed to an integral over $\omega$ by using the Poisson resummation formula
	\begin{align}
		I&=\sum_l\int \frac{dk}{2\pi} \frac{d\omega}{2\pi}\frac{e^{-i l\beta\omega}}{\omega^2+k^2+m^2}\non
		&= I_0+\int \frac{dk}{2\pi}\frac{n_k(\beta)}{E_k}.\label{tadpole integral}
	\end{align}
	The $l=0$ term is the same as the zero-temperature tadpole integral, and this is referred to as $I_0$. In the $l\neq 0$ terms, the $\omega$ integral has been performed, and the geometric series summed.
	
	Now noting that the mass counterterm at this order is $\delta m^2 = -\frac{\lambda}{2}I_0,$ it is seen that all appearances of $I_0$ cancel in \eqref{eq free energy 1st order}	\begin{align}
		f^{(1)}&=\frac{m^2 g^2}{8}\left(\int \frac{dk}{2\pi}\frac{n_k(\beta)}{E_k}\right)^2\non
		&= \frac{1}{L} \,\frac{1}{2!}\int \frac{d k_1}{2\pi}\frac{1}{2E_1}\frac{d k_2}{2\pi}\frac{1}{2E_2}\frac{e^{-\beta(E_1+E_2)}}{\left(1-e^{-\beta E_1}\right)\left(1-e^{-\beta E_2}\right)} m^2g^2 L.\label{f first order phi4}
	\end{align}
If we momentarily ignore the two factors of $(1-e^{-\beta E})^{-1}$, this is clearly in the form of the equation \eqref{Eq DMB formula single T},
	\begin{align*}
		f^{(1)}= \frac{1}{L}\int d\alpha \,e^{-\beta E_\alpha}T^{(1)}_{\alpha\alpha},
	\end{align*}
	where $\alpha$ represents a state of two identical relativistic particles with momenta $k_1$ and $k_2$, and $d\alpha$ is the appropriate measure for a trace over these states. The first-order forward scattering $T$ matrix element is just the vertex factor $m^2 g^2$ times the volume $L$.
	
		Diagrams involving a single higher-order vertex, such as the second-order diagram involving the $\phi^6$ vertex in Fig \ref{Fig first order}, follow similarly. The counterterm diagrams will again end up canceling the $I_0$ terms in the tadpole integrals, and the net contribution to the free energy density is
		 	\begin{align}
		 	f_{\phi^{2n}}= \frac{1}{L} \,\frac{1}{n!}\int \left(\prod_{i=1}^n\frac{d k_i}{2\pi}\frac{1}{2E_i}\frac{e^{-\beta E_i}}{1-e^{-\beta E_i}}\right) m^2g^{2n-2} L.\label{f n-1 order}
		 \end{align}
	 		
\begin{figure}
		\centering
		\includegraphics[width=0.15\textwidth]{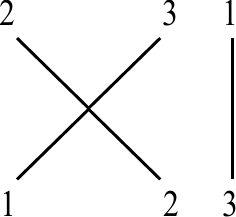}
		\caption{An example of a `disconnected' forward scattering amplitude that nevertheless contributes to the free energy due to identical particle exchange. The labels at the bottom refer to incoming momenta, and the top outgoing momenta.}\label{Fig exchange example}
	\end{figure}

	This has the appropriate measure for $n$ relativistic particles, and involves the $n$-body forward scattering amplitude $T_{\alpha\alpha}=m^2g^{2n-2} L$, but once again there are additional factors of  $(1-e^{-\beta E_i})^{-1}$. These factors have to do with bosonic statistics, and they arise by considering auxiliary particles in the states $\alpha$ which do not participate in the two-body scattering. Due to identical particle exchange there will be parts of the forward scattering $T$ matrix which are still connected in the sense that they involve a single factor of $L$ and can not be written as a product of two or more free energy terms. An example is given in Fig \ref{Fig exchange example}. The auxiliary particle line leads to a delta function $2E_3 2\pi\delta(k_3-k_1)$ that cancels the trace over the third particle momentum and leads to a correction which agrees with the first non-trivial term in the expansion of $(1-e^{-\beta E_1})^{-1}$.
	
These $(1-e^{-\beta E_i})^{-1}$ factors involve arbitrarily high particle number $N$ as far as the fugacity expansion \eqref{Eq DMB formula grand canonical} is concerned, but each term in the expansion involves only some finite number of $n$ particles interacting through non-trivial scattering. It is useful to consider an expansion in the number of non-trivially interacting particles, which will be referred to as the \emph{n-body} expansion. An ``$n$-body correction'' like \eqref{f n-1 order} will generically involve $n$ factors of bosonic occupation numbers $n_i$.

For later reference we will write the three-body correction due to the $\phi^6$ vertex explicitly in terms of occupation numbers as a function of rapidity
\begin{align}
		f_{\phi^6}&=\frac{1}{3}\frac{m^2 g^4}{16}\left(\int \frac{d\theta}{2\pi}n(\theta)\right)^3,\label{f phi6 3-body}
	\end{align}
	where $n(\theta)=\left(e^{\beta m\cosh\theta}-1\right)^{-1}$.

	%
	
	\subsection{Cutting a general diagram}\label{Sec Intro diagrams general cutting}

	The previous example involved using the Poisson resummation formula to write finite temperature propagators in terms of an infinite sum over an index $l$. The $l=0$ term which is present at zero-temperature was split from the rest. Now we will generalize this procedure.
	
	A general $r$-loop free energy diagram $\delta f$ may be written in terms of $r$ independent frequencies, each with its own summation over exponentials due to the Poisson resummation formula,
	\begin{align*}
		\delta f =  \int \prod_{j=1}^r\frac{dk_j}{2\pi}\frac{d\omega_j}{2\pi}\frac{\sum_{l_j}e^{-il_j\beta \omega_j}}{\omega_j^2 + E_j^2}\mathcal{M}\left(\omega, k\right).
	\end{align*} 
	The factor $\mathcal{M}$ collects all of the combinatorial and vertex factors, as well as any additional propagators which are fixed by the delta functions at the vertices to be functions of the $r$ independent $\omega_j, k_j$.
	
	The summation over $l_j$ may be broken up into the $l_j=0$ term and the $l_j\neq 0$ terms,
	\begin{align*}
		\sum_{l_j} e^{-i l_j \beta \omega_j}= 1+ \sum_{\sigma_j=\pm 1}\sum_{l_j>0} e^{-i \sigma_j l_j \beta \omega_j}.
	\end{align*}
	After decomposing the $r$ propagators into two terms in this manner, there are a total of $2^r$ terms in $\delta f$. We refer to the propagators that keep the $l\neq 0$ terms as \emph{cut}, and the those that keep the $l=0$ term as \emph{uncut}.  We may index the $2^r$ terms in $\delta f$ by a set $\pi$ which contains the indices of the cut propagators. The other uncut propagators are integrated over and absorbed into the $\mathcal{M}$ factor, which is now denoted $\mathcal{M}_\pi$.
	\begin{align*}
		\delta f =  \sum_\pi\int \prod_{j\in \pi}\frac{dk_j}{2\pi}\frac{d\omega_j}{2\pi}\sum_{\sigma_j=\pm 1}\sum_{l_j>0} \frac{e^{-i\sigma_j l_j\beta \omega_j}}{\omega_j^2 + E_j^2}\mathcal{M}_\pi\left(\omega, k\right).
	\end{align*} 
	The quantity $\mathcal{M}_\pi$ is easy to understand diagramatically. It is just given by cutting the cut propagators of the original free energy diagram and taking them to be external legs. The remaining uncut propagators of the diagram are taken to be at zero-temperature. $\mathcal{M}_\pi$ is an off-shell forward scattering amplitude depending on the frequency and spatial momenta of the cut lines.
	
	If we momentarily assume that $\mathcal{M}_\pi(\omega)$ is analytic then only the pole at $\omega_j = -i\sigma_j E_j$ due to the cut propagator will contribute to the $\omega_j$ integral. Then after integration over $\omega$ and summation of the geometric series in $l$ we get the formula,
	\begin{align*}
		\delta f =  \sum_\pi \int \prod_{j\in \pi}\frac{dk_j}{2\pi}\frac{n(E_j)}{2E_j}\sum_{\sigma} \mathcal{M}_\pi\left(-i\sigma E, k\right).
	\end{align*} 
	
	This is now roughly in the form of the DMB formula \eqref{Eq DMB formula single T}. $\mathcal{M}_\pi$ has been put on-shell, and the signs $\sigma$ determine whether the momentum of a cut line is considered to be incoming or outgoing. If there are $n$ cut lines belonging to $\pi$ then this is an $n$-body contribution to the free energy diagram $\delta f$, and the factors of $n(E)$ automatically contain the higher-order corrections due to particle exchange.
	
	Generically $\mathcal{M}_\pi(\omega)$ will not be analytic. There will be an additional contributions to the frequency integral from poles or branch cuts in $\mathcal{M}_\pi$. This will be easy enough to deal with in Sec \ref{Sec On shell} for the case of diagrams with only on-shell contributions. The result will agree with the full DMB formula that includes the higher order in $\hat{T}$ corrections which are neglected in \eqref{Eq DMB formula single T}.
	\subsection{The bubble and melon diagrams}\label{Sec Intro diagrams bubble melon}
	
\begin{figure}
		\centering
		\includegraphics[width=0.5\textwidth]{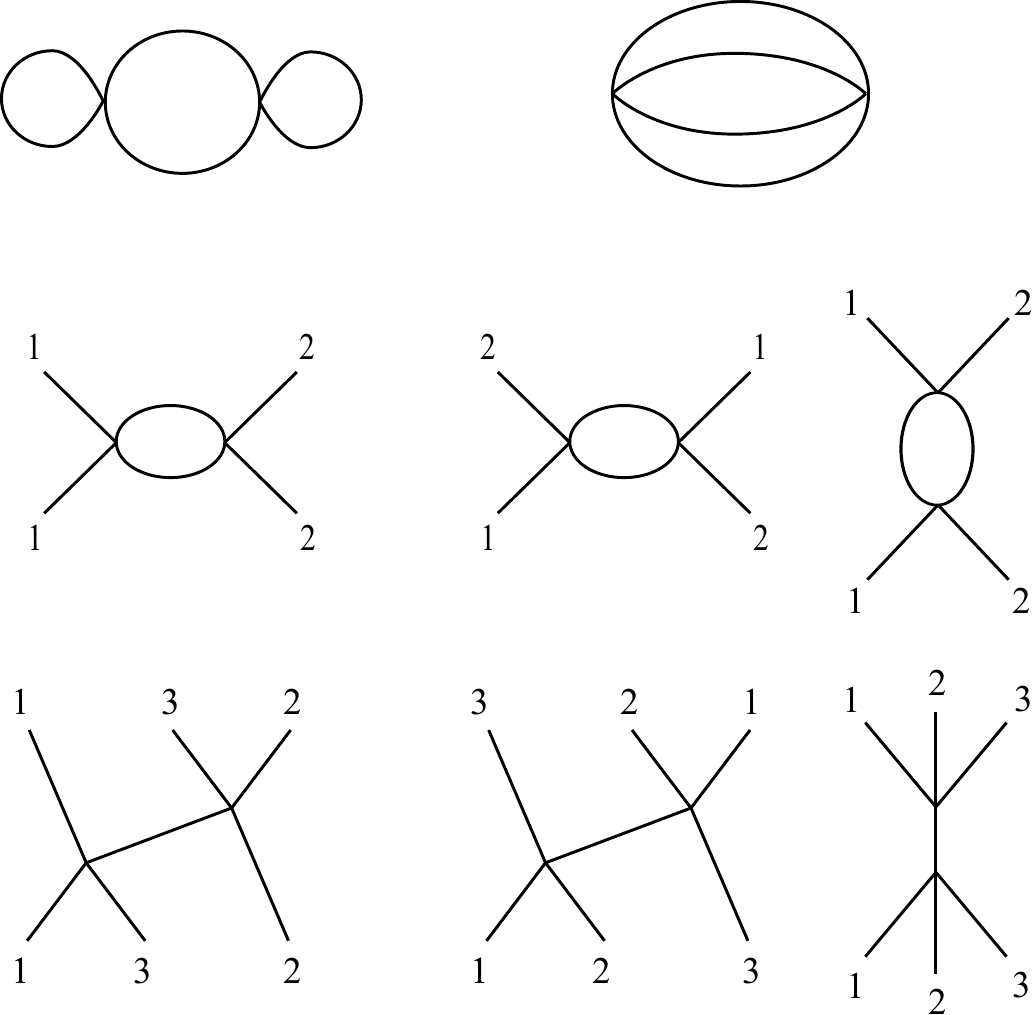}
		\caption{The `bubble' and `melon' corrections to the free energy of a $\phi^4$ theory at second order, on the left and right of the top row, respectively. The two-body and three-body forward scattering amplitudes corresponding to these diagrams are shown in the second and third rows, with numbers to indicate distinct momenta.}\label{Fig bubble melon}
	\end{figure}

	For now let us consider the free energy diagrams at second-order in the sinh-Gordon and $\phi^4$ theories. As discussed in Sec \ref{Sec Intro DMB mult part}, perturbation theory at this order will clarify how three-particle scattering amplitudes are treated in the DMB approach. The diagram involving a single sextic vertex in sinh-Gordon has already been considered in \eqref{f phi6 3-body}. The remaining two diagrams ---the \emph{bubble} and \emph{melon}--- are shown in Fig \ref{Fig bubble melon}.
	
	The previous section has made clear which forward scattering amplitudes are associated to a given thermal field theory diagram. Lines of the thermal field theory diagram are cut to create incoming and outgoing external lines with the same momenta. Conversely, we may `glue together' incoming and outgoing lines with the same momentum in a forward scattering amplitude (the lines labeled by the same number in Fig \ref{Fig bubble melon}) to recover a free energy diagram.
	\subsubsection{The bubble diagram}\label{Sec Intro diagrams bubble}
	The bubble diagram is shown on the left side of Fig \ref{Fig bubble melon}. As in the single vertex case, free energy diagrams involving counterterms will cancel the uncut $I_0$ part of any tadpoles, so the two tadpoles on the sides of the diagram are always cut.
	
	The two-body part of the diagram $f_{\text{bubble},\, 2}$ involves an uncut central bubble. Since in this case the amplitude $\mathcal{M}_\pi$ does not depend on the cut momenta, it can be readily calculated
\begin{align}
f_{\text{bubble},\,2}&=-\frac{m^4 g^4}{16}\left(\int \frac{d\theta}{2\pi}n(\theta)\right)^2\frac{1}{4\pi m^2}.\label{f bubble 2-body}
\end{align}
	
	The scattering amplitude that would correspond to the three-body part of the bubble diagram $f_{\text{bubble},\,3}$ is given on the left of the bottom row of Fig \ref{Fig bubble melon}. This diagram has a divergence due to the internal propagator being fixed by momentum-conservation to have the on-shell momentum labeled $3$. This is an example of an `on-shell contribution.'
	
	Going back to the general cutting argument, this divergence just indicates that $\mathcal{M}_\pi$ also has a pole as a function of $\omega_3$, and this needs to be taken into account in the integration over $\omega_3$. So dealing with the on-shell contribution is no major difficulty, it just involves integrating over a double pole
	\begin{align}
	 f_{\text{bubble},\,3}&=-\frac{m^4 g^4}{16}\left(\int \frac{d\theta}{2\pi}n(\theta)\right)^2	\int\frac{dk_3}{2\pi}\sum_{\sigma_3,l_3}\int\frac{d\omega_3}{2\pi}\frac{ e^{-i\sigma_3 l_3 \beta \omega_3}}{\left(\omega_3^2+E_3^2\right)^2}\non&=-\frac{m^2 g^4}{16}\left(\int \frac{d\theta}{2\pi}n(\theta)\right)^2	\int\frac{dk_3}{2\pi}\frac{m^2}{2E_3^2}\left(-\frac{d}{dE_3}+\frac{1}{E_3}\right)n(E_3).\label{f bubble 3-body 1}
	\end{align}

	This expression will be derived from the DMB formula in Sec \ref{Sec On shell}. By using similar manipulations to those in Appendix \ref{Sec Appendix large N xS} it can be brought to a simplified form
		\begin{align}
		f_{\text{bubble},\,3}&=\frac{1}{3!}\frac{m^2 g^4}{16}\beta \der{}{\beta}\left(\int \frac{d\theta}{2\pi}n(\theta)\right)^3.\label{f bubble 3-body}
	\end{align}
	
	
	\subsubsection{The melon diagram: two-body part}\label{Sec Intro diagrams melon 2body}

	The forward scattering diagrams corresponding to cutting the melon diagram are shown on the right of Fig \ref{Fig bubble melon}. In the following sections we will use these diagrams in the DMB approach to evaluate the melon diagram and show it involves only contributions with collinear momenta. Of course the melon diagram may also be evaluated directly from thermal field theory, and this is shown using methods similar to \cite{ bugrijShadura1995, massiveBasketball} in Appendix \ref{Sec Appendix melon cut}.
	
	Note that the two-body amplitudes can be written in terms of the one-loop integral
		\begin{align}
		J(p)&\equiv \int \frac{d^2 q}{(2\pi)^2}\frac{1}{q^2+m^2}\frac{1}{(p-q)^2+m^2}.
	\end{align}
The bubble diagram was associated to the u-channel of two-body scattering, with no net momentum flowing through the loop, and \eqref{f bubble 2-body} indeed involves the factor $J(0)=(4\pi m^2)^{-1}$. The melon diagram instead involves the t-channel and s-channel. By the DMB approach, the two-body part of the melon should be
	\begin{align}
		f_{\text{melon,\,2}}=-\frac{m^4g^4}{2}\frac{1}{2!}\int \left(\prod_{j=1,2} \frac{dk_j}{2\pi}\frac{n_j}{2E_j}\right) \left[J(-i(E_1-E_2))+\mathcal{P}J(-i(E_1+E_2))\right].\label{f melon 2-body J}
	\end{align}
	The argument of the s-channel term $J(-i(E_1+E_2))$ sits on a branch cut, and some sort of $i\epsilon$ prescription is needed to define it. The principal value prescription arising from the second-order DMB formula \eqref{DMB principal value} is the correct one.
	
This can be calculated further, using the expression
\begin{align*}
	J(p)=\frac{1}{2\pi p^2 \xi}\log \left(\frac{\xi+1}{\xi-1}\right),\qquad \xi\equiv \sqrt{1+4\frac{m^2}{p^2}}.
\end{align*}
The sum of terms in \eqref{f melon 2-body J} is proportional to a delta function,
\begin{align*}
	&J\left(-i(E_1-E_2)\right)+\frac{1}{2}\left[J(-i(E_1+E_2)+\epsilon)+J(-i(E_1+E_2)-\epsilon)\right]\non&\qquad=	\frac{\log e^\theta}{4\pi m^2 \sinh\theta}+\frac{1}{2}\left[-\frac{\log \left(-e^\theta+i\epsilon\right)}{4\pi m^2 \sqrt{\sinh^2 \theta-i\epsilon}}-\frac{\log \left(-e^\theta-i\epsilon\right)}{4\pi m^2 \sqrt{\sinh^2 \theta+i\epsilon}}\right]\non
	&\qquad=\frac{2\pi }{16 m^2}\left[\frac{i}{ \pi\sqrt{\sinh^2 \theta+i\epsilon}}-\frac{i}{\pi\sqrt{\sinh^2 \theta-i\epsilon}}\right]\non
	&\qquad=\frac{2\pi}{16 m^2}\delta(\theta).
\end{align*}
So in total
\begin{align}
	f_{\text{melon,\,2}}=-\frac{1}{16}\frac{m^2g^4}{16}\int \frac{d\theta}{2\pi}n(\theta)^2.\label{f melon 2-body}
\end{align}

	\subsubsection{The melon diagram: three-body part}\label{Sec Intro diagrams melon 3body}
	
	The three-body part of the melon diagram is similarly calculated from the three-body amplitudes in Fig \ref{Fig bubble melon},
	\begin{align}
f_{\text{melon,\,3}}=-m^4 g^4\frac{1}{3!}\int \left(\prod_{j=1,2,3} \frac{dk_j}{2\pi}\frac{n_j}{2E_j}\right)&\left[\mathcal{P}\frac{3}{-\left(E_1+E_2-E_3\right)^2+\left(k_1+k_2-k_3\right)^2+m^2}\right.\non
&\qquad	\left.+\frac{1}{-\left(E_1+E_2+E_3\right)^2+\left(k_1+k_2+k_3\right)^2+m^2}\right].\label{f melon 3-body prop}
	\end{align}
As long as the momenta are distinct the propagators are not on-shell, and the principal value regularization may be dropped.

The first propagator in \eqref{f melon 3-body prop} may be written in terms of rapidity as,
	\begin{gather}
		\frac{1}{-\left(E_1+E_2-E_3\right)^2+\left(k_1+k_2-k_3\right)^2+m^2}=\frac{e^{\theta_1+\theta_2+\theta_3}}{m^2\left(e^{\theta_1}-e^{\theta_3}\right)\left(e^{\theta_2}-e^{\theta_3}\right)\left(e^{\theta_1}+e^{\theta_2}\right)}.\label{naive melon three-body}
	\end{gather}
There are two additional diagrams where index 3 is swapped with index 1 or 2 (hence the factor of three in \eqref{f melon 3-body prop}). Adding these terms with permuted indices leads to
\begin{align}
\frac{e^{\theta_1+\theta_2+\theta_3}}{m^2\left(e^{\theta_1}-e^{\theta_3}\right)\left(e^{\theta_2}-e^{\theta_3}\right)\left(e^{\theta_1}+e^{\theta_2}\right)}+\left(3\leftrightarrow 1\right)+\left(2\leftrightarrow 3\right) = \frac{e^{\theta_1+\theta_2+\theta_3}}{m^2\left(e^{\theta_1}+e^{\theta_3}\right)\left(e^{\theta_2}+e^{\theta_3}\right)\left(e^{\theta_1}+e^{\theta_2}\right)}.
\end{align}
	But this simply cancels with the second propagator in \eqref{f melon 3-body prop},
	\begin{gather}
		\frac{1}{-\left(E_1+E_2+E_3\right)^2+\left(k_1+k_2+k_3\right)^2+m^2}=-\frac{e^{\theta_1+\theta_2+\theta_3}}{m^2\left(e^{\theta_1}+e^{\theta_3}\right)\left(e^{\theta_2}+e^{\theta_3}\right)\left(e^{\theta_1}+e^{\theta_2}\right)}.\label{melon 3-body naive}
	\end{gather}
	
	The cancelation of the three-body forward scattering diagrams associated to the melon diagram is not surprising given the cancelations which must take place in the integrable sinh-Gordon model.\footnote{It can also be shown that the three-body diagrams associated with the bubble, displaced slightly from the forward direction, are canceled entirely by the three-body diagram given by the $\phi^6$ vertex.} But actually the three-body part of the melon diagram does not vanish. The principal value regulation may not be ignored when the momentum become collinear, and a calculation in Appendix \ref{Sec Appendix melon 3body} shows that the integrand of \eqref{f melon 3-body prop} is proportional to a delta function, much as in the two-body case. The final result is
	\begin{align}
		f_{\text{melon,\,3}}
&	=-\frac{1}{24}\frac{m^2 g^4}{16}\int \frac{d\theta}{2\pi}n(\theta)^3.	\label{f melon 3-body}
	\end{align}

	 In Sec \ref{Sec TBA} these same results will be derived from the TBA. It will be discussed why similar collinear contributions are expected at higher order in perturbation theory, and how they agree with the appearance of bound states upon switching the sign of the coupling in the theory.
	
	\section{On-shell contributions and large $N$ theories}\label{Sec On shell}
	The bubble diagram is part of a broader class of diagrams where the integrals over distinct loop momenta factorize. These are the same diagrams which would be singled out at leading order in a large $N$ expansion. To study the net effect of these diagrams we consider extending the field in the $\phi^4$ theory \eqref{Lagr phi4} to an $N$ component field $\phi$ with Lagrangian
	\begin{align*}
	\Lagr=\frac{1}{2}\left(\partial \phi \cdot \partial \phi + \left(m^2+\delta m^2\right) \phi^2\right)+\frac{\lambda}{4N}\left(\phi^2\right)^2.
\end{align*}
	The counterterm $\delta m^2 = -\lambda I_0$ is chosen to ensure that $m^2$ is the physical mass at large $N$ and zero temperature. The quartic term may be simplified by introducing a Hubbard-Stratonovich field $\alpha$
	\begin{align}
		\Lagr=\frac{1}{2}\left(\partial \phi \cdot \partial \phi + m^2 \phi^2\right)-\frac{N}{\lambda}\alpha^2+\alpha\left(\phi^2-N I_0\right).\label{Lagr large N}
	\end{align}
Here $\alpha$ has a quadratic term that vanishes in the strong coupling limit $\lambda\rightarrow \infty$. In this limit $\alpha$ can be understood as a Lagrange multiplier field, and the model reduces to the integrable $O(N)$ non-linear sigma model.

Note that the counterterm was set to ensure that $\alpha$ has no expectation value at leading order for zero-temperature, but at finite temperature it does pick up an expectation value. We may absorb the expectation value into a temperature-dependent mass $M(\beta)$, and a redefined field $\alpha_\beta \equiv \alpha-\frac{1}{2}(M^2-m^2)$,
\begin{align}
	\Lagr=\frac{1}{2}\left(\partial \phi \cdot \partial \phi + M(\beta)^2 \phi^2\right)+\alpha_\beta\left(\phi^2-N I_0\right)-\frac{1}{2}\left(M(\beta)^2-m^2\right)N I_0.\label{Lagr large N Mbeta}
\end{align}

The leading-order free energy of this form of the $O(N)$ non-linear sigma model was considered by Balog and Hegedus in \cite{balogHegedusVirial2001}. They used an exact expression for the free energy density which follows directly from \eqref{Lagr large N Mbeta}, 
\begin{align}
	f=\frac{Nm^2}{2\pi}\left[-\frac{M^2}{m^2}\sum_{j=1}^\infty K_2\left(jm\beta \frac{M}{m}\right) +\frac{1}{4}\left(\frac{M^2}{m^2}-1\right)\right].\label{Eq BH f}
\end{align}
This may be expanded in orders of fugacity by solving the gap equation \cite{luscher1982} for $M/m$ order-by-order,
\begin{align}
	\log \frac{M}{m}=2\sum_{j=1}K_0\left(j\beta m \frac{M}{m}\right).\label{Eq BH gap}
\end{align}

Using this method, Balog and Hegedus were able to find the second-order in fugacity contribution $f_{(2)}$ and show it agrees with the DMB formula for two-particle scattering. In principle there is no obstacle to using this method to calculate arbitrary orders $f_{(j)}$, but matching to the DMB even for $f_{(3)}$ encounters the problem of the forward scattering singularities. In the following sections we will give an alternate method to calculate the free energy density, and show how it may be matched to forward scattering amplitudes at arbitrary order.

However, note in passing that there is a direct connection between the thermal mass $M(\beta)$ and forward scattering amplitudes along the same lines as the DMB relation. In \cite{luscher1984,luscher1986 I}, L\"{u}scher derives a relation which in our present circumstances takes the form
\begin{align}
	\frac{M-m}{m}=-\frac{1}{2m^2}\int \frac{dp}{2\pi}\frac{1}{2E}e^{-\beta E}F +\order{e^{-2m\beta}}.
\end{align}
Here $F$ is the sum of the two-body forward scattering amplitudes with the other particles of the theory, which works out to be $-8\pi m^2$ in the case of the large $N$ limit of the $O(N)$ non-linear sigma model. Expanding the gap equation \eqref{Eq BH gap} to first order in fugacity we recover this result. Higher order in fugacity corrections to $M^2$ can be derived from thermal self-energy diagrams calculated using the Feynman rules of the following section, and these correspond to multi-particle  forward scattering amplitudes in the same way.

\subsection{Diagrams in the large $N$ theory}\label{Sec On shell diagrams}

The equations \eqref{Eq BH f} and \eqref{Eq BH gap} give a compact expression for the free energy, but actually solving them order-by-order quickly leads to a tedious calculation. We will organize the calculation of the free energy by an approach involving thermal field theory diagrams, which then can be quickly related to forward scattering amplitudes through the cutting rules discussed in Sec \ref{Sec Intro diagrams general cutting}.

Let us return to the form of the Lagrangian involving the zero-temperature mass $m^2$ \eqref{Lagr large N}. We may keep finite $\lambda$ dependence. There is a free boson contribution $f_1$ coming from the quadratic action of the $\phi$ field
\begin{align}
	f_1 = \frac{N}{2}\int_\beta \frac{d^2p}{(2\pi)^2} \log\left(p^2+m^2\right)=-\frac{Nm^2}{\pi}\sum_{j=1}^\infty  \frac{K_1(j\beta m)}{j\beta m}.\label{eq free free energy}
\end{align}
There are further corrections coming from thermal field theory diagrams involving internal $\alpha$ propagators. Since each $\alpha$ propagator introduces a factor of $1/N$, at leading order it must be paired with a $\phi$ loop, which introduces a compensating factor of $N$. This implies that at leading order there will be no momentum flowing through the $\alpha$ propagators.

So to calculate the leading order diagrams, we must consider a class of integrals which may be derived from the basic tadpole integral $I$ in \eqref{tadpole integral},
	\begin{align}
		I_{,n}\equiv 	  \int_\beta \frac{d^2p}{(2\pi)^2} \frac{1}{\left(p^2+m^2\right)^{n+1}}=\frac{(-1)^n}{n!}\frac{d^n I}{d(m^2)^n} .
	\end{align}
These integrals may be calculated by taking derivatives of the formula,
\begin{align}
	I= \frac{1}{4\pi}\log \frac{M_{UV}^2}{m^2}+\frac{1}{\pi}\sum_{j=1}^\infty K_0(j\beta m),
\end{align}
where $M_{UV}$ is a UV cutoff. Using the Bessel function identity $K_{\alpha+1}(x) =\left( \frac{\alpha}{x}-\derOrd{}{x}\right)K_\alpha(x)$, we can show
\begin{gather}
	I_{,n}=\frac{1}{4\pi n m^{2n}}\left[1+\frac{x_n}{2^{n-2}(n-1)!}\right],\\
	x_n\equiv \sum_{j=1}^\infty \left(j\beta m\right)^n K_n(j\beta m).\label{Def xS}
\end{gather}
The integrals $I_{,n}$ neatly decompose into an `uncut' piece $I_{0,n}=(4\pi n m^{2n})^{-1}$, and a `cut' piece involving the quantity $x_n$. The uncut bubbles $I_{0,1}$ will be used to form effective propagators for $\alpha$.

An effective action for the zero mode of the $\alpha$ field may be found by integrating out the $\phi$ field in \eqref{Lagr large N},
\begin{align}
	\Lagr_{\text{eff}} = -N\left(\lambda^{-1}+I_{,1}\right)\alpha^2 + N\sum_{j=3}^\infty \frac{(-2)^{j-1}}{j}I_{,j-1}\alpha^j.
\end{align} 
The effective propagator may be written in terms of a geometric series of cut bubbles,
\begin{gather}
	-\frac{\lambda}{2N}\frac{1}{1+\lambda I_{,1}}=-\frac{2\pi m^2 \Lambda}{N}\sum_{j=0}^\infty \left(-\frac{2\pi m^2\Lambda}{N} \frac{N x_1}{\pi m^2}\right)^j,\qquad\Lambda \equiv \frac{\lambda}{\lambda+4\pi m^2}.
\end{gather}
The renormalized coupling $\Lambda$ is simply equal to $1$ in the non-linear sigma model limit.\\

\noindent Now we may present the Feynman rules for large $N$ free energy diagrams:
\begin{itemize}
	\item $\alpha$ propagators are drawn as dashed lines, each contributing $-2\pi m^2 \Lambda/N$.	
	\item All diagrams are tree diagrams in terms of the $\alpha$ propagators.
	\item $\phi$ loops are drawn as solid lines, and are always taken to be `cut.'
	\item An uncut $\alpha^j$ vertex is drawn as a solid point, and contributes a factor $N(-2)^{j-1}I_{0,j-1}/j$.
	\item A cut $\alpha^j$ vertex is drawn as a $\phi$ loop, and contributes an additional factor of $2^{3-j}x_{j-1}/(j-2)!$.
	\item A $\phi$ tadpole contributes $N x_0/\pi$. 
	\item $\phi$ bubbles may be inserted in an $\alpha$ propagator, contributing factors of $N x_1/(\pi m^2)$.
\end{itemize}

\begin{figure}
	\centering
	\includegraphics[width=0.65\textwidth]{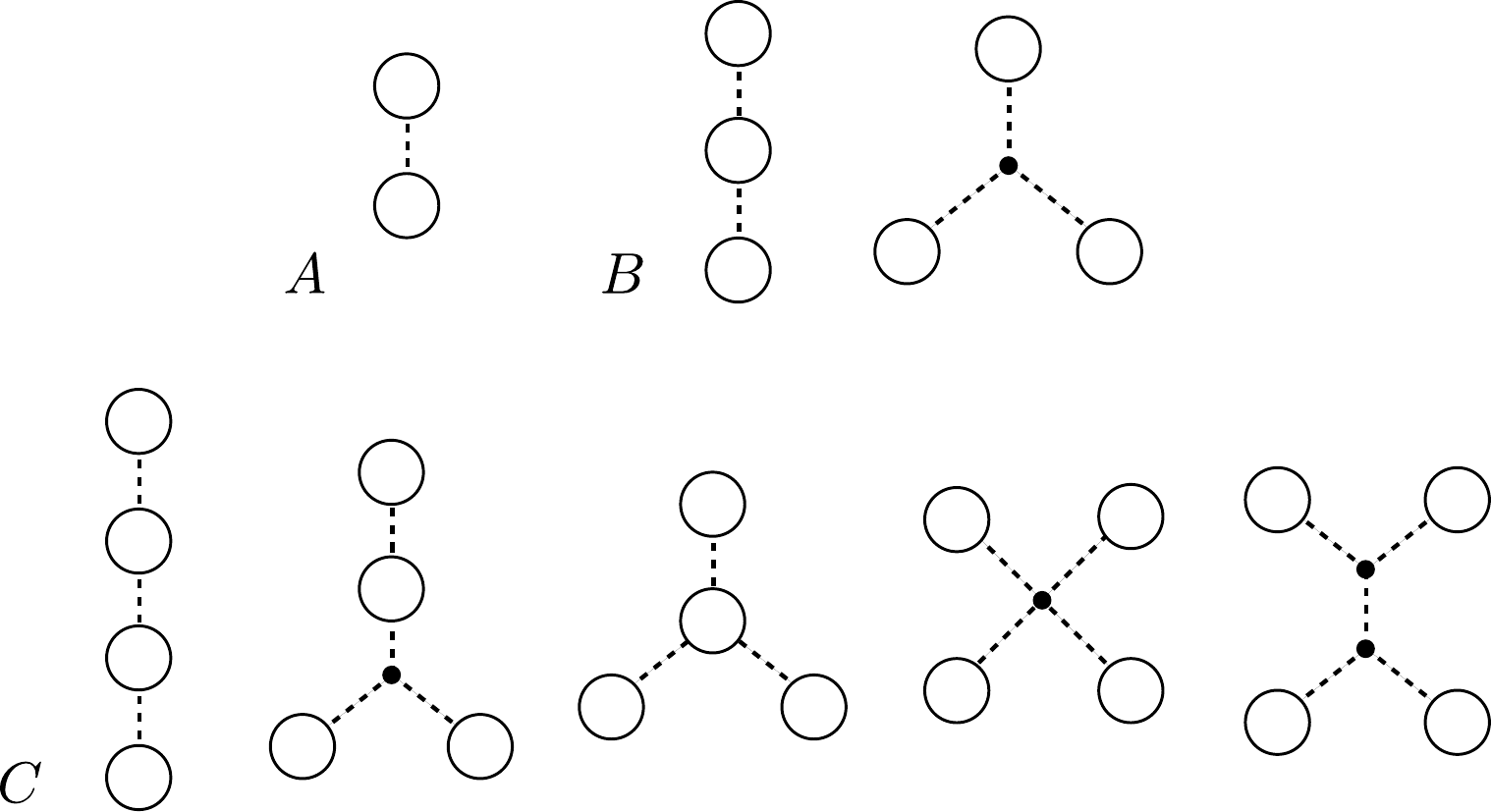}
	\caption{Large $N$ free energy diagrams up to four $\phi$ loops, corresponding to two-body (A), three-body (B), and four-body (C) scattering amplitudes.}\label{Fig ONfreeEnergy}
\end{figure}

The free energy diagrams up to four $\phi$ loops are shown in Fig \ref{Fig ONfreeEnergy}. Together with the `one-body' contribution \eqref{eq free free energy}, they may be expanded up to fourth-order in fugacity to match with the calculation of $f_{(4)}$ via the gap equation \eqref{Eq BH gap}.

\subsection{On-shell contributions from the DMB formula}\label{Sec On shell DMB}

In general, a free energy diagram with $n$ $\phi$ loops corresponds to the $n$-body forward scattering amplitude where each loop is cut. Diagrams that only involve $\phi$ tadpoles are straightforward to match to the DMB formula, as in Sec \ref{Sec Intro diagrams single vertex}. However, a bubble insertion in an $\alpha$ propagator will correspond to a single internal propagator that goes on-shell, much like the three-body part of the bubble diagram in Sec \ref{Sec Intro diagrams bubble}. And a $\phi$ loop vertex with $p$ attached $\alpha$ propagators will involve $p-1$ internal propagators going on-shell.

These are all what we have been referring to as on-shell contributions, and they are not specific to the large $N$ theory. From the previous section it is clear that a forward scattering diagram with $j-1$ internal propagators going on-shell should involve a factor of $x_{j-1}$, which is a sum of Bessel functions $K_{j-1}$. So the problem of dealing with on-shell contributions in the DMB approach amounts to deriving this sum of Bessel functions from the DMB formula.

To do this it is helpful to use another expression for $x_s$, which is derived in Appendix \ref{Sec Appendix large N xS},
\begin{align}
		x_s = \int \frac{dk}{2E}\frac{m^{2s}}{E^s}\sum_{l=0}^s \frac{(s+l)!}{(2E)^l(s-l)!l!}\left(-\frac{d}{dE}\right)^{s-l}n(E).\label{eq xS expansion}
\end{align}
In particular, the simplest case is
\begin{align}
	x_1 = \int dk\frac{ m^2}{2E^2}\left(-\frac{d}{dE}+\frac{1}{E}\right)n(E),\label{eq x1 expansion}
\end{align}
which is clearly involved in the three-body part of the bubble diagram \eqref{f bubble 3-body 1}. We will first sketch how the bubble diagram is derived from the DMB formula, and then move on to the general case.

For simplicity we will ignore particle exchange corrections in the discussion that follows. This amounts to truncating the summation in $x_s$ to the first term and replacing $n(E)$ by $e^{-\beta E}$. For example \eqref{eq x1 expansion} becomes
\begin{align}
\beta m\,	K_1(\beta m)=\int dk\frac{ m^2}{2E^2}\left(\beta+\frac{1}{E}\right)e^{-\beta E}.\label{eq x1 no exchange}
\end{align}

\subsubsection{DMB calculation of the bubble diagram}\label{Sec On shell DMB bubble}

\begin{figure}
	\centering
	\includegraphics[width=0.4\textwidth]{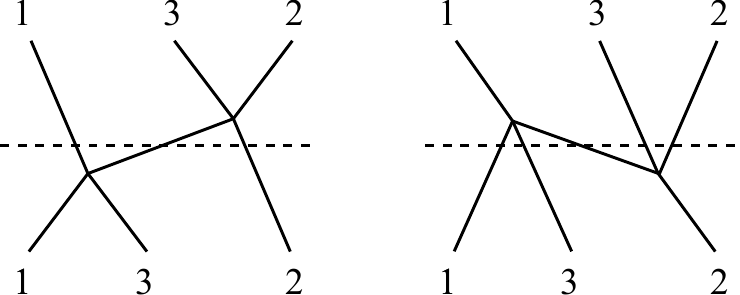}
	\caption{Two distinct intermediate states contribute to the three-body part of the bubble diagram. The particles that cross the dotted line are acted upon by the propagator $\mathcal{P}G_0$. }\label{Fig OFPT}
\end{figure}

Given the second-order DMB formula \eqref{DMB principal value}, the bubble diagram will involve the amplitude $\langle \alpha| (G_0-\bar{G}_0)\,V \mathcal{P} G_0 V|\alpha\rangle$. The three-particle state is $|\alpha\rangle=a^\dagger(p_1)a^\dagger(p_2)a^\dagger(p_3)|0\rangle$ and the interaction $V=\frac{m^2g^2}{4!}\int d^1x\phi(x)^4$. This involves the free field operator
\begin{align}
	\phi(x)=\int \frac{dk}{2\pi}\frac{1}{\sqrt{2E_k}}\left(a^\dagger(k)e^{ikx}+a(k)e^{-ikx}\right).
\end{align}
This is essentially an old-fashioned perturbation theory calculation, and the rightmost copy of $V$ can act on the state $\alpha$ in two different ways leading to an intermediate state with either three or five particles, see Fig \ref{Fig OFPT}. The diagram with a three-particle intermediate state is the source of the forward scattering divergence, and it involves a propagator with the same energy difference $E-E_\alpha$ that appears in the would-be delta function $G_0-\bar{G}_0$.

As noted already in \cite{dmb}, this is equivalent to a derivative of the energy delta function
\begin{align*}
	\left(G_0-\bar{G}_0\right)\mathcal{P}G_0 &= \frac{-2i\epsilon }{(E-E_\alpha)^2+\epsilon^2}\frac{E-E_\alpha}{(E-E_\alpha)^2+\epsilon^2}=-\frac{1}{2}\der{}{E}\left(G_0-\bar{G}_0\right).
\end{align*}
The $E$ derivative will produce a factor of $\beta$ by acting on $e^{-\beta E}$ after integration by parts, so effectively the propagator is replaced by $$\mathcal{P}G_0\rightarrow -\frac{1}{2}\beta.$$

On the other hand, in the diagram with a five-particle intermediate state, all of the combinatorics end up being the same but the propagator never goes on-shell so we can safely set $E\rightarrow E_1+E_2+E_3,$
$$\mathcal{P}G_0\rightarrow \mathcal{P}\frac{1}{E-(E_1+E_2+3E_3)}= -\frac{1}{2}\frac{1}{E_3}.$$
This is how the factor $\beta+1/E_3$ in \eqref{eq x1 no exchange} arises from the DMB perspective.

\subsubsection{Higher-order on-shell contributions}\label{Sec On shell DMB higher}

\begin{figure}
	\centering
	\includegraphics[width=0.6\textwidth]{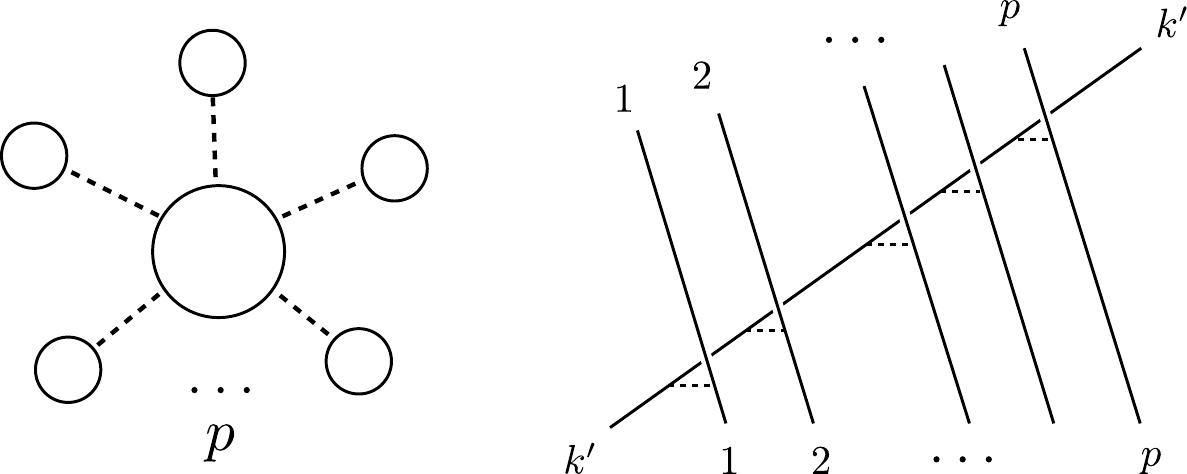}
	\caption{The free energy diagram $f_{\alpha^p}$ (left). This corresponds to a forward scattering amplitude (right) involving $p+1$ particles.}
	\label{Fig VertexDiagrams}
\end{figure}

The basic large $N$ free energy diagram involving a single cut $\alpha^p$ vertex and $p$ tadpoles is shown in Fig \ref{Fig VertexDiagrams}. It corresponds to a forward scattering amplitude among $p+1$ particles, where $p-1$ internal propagators all go on-shell with momentum $k'$. Following the Feynman rules in Sec \ref{Sec On shell diagrams}, the free energy $f_{\alpha^p}$ associated to this diagram is
\begin{align}
	f_{\alpha^p}=-\frac{Nm^2}{\pi}\frac{(-2\Lambda)^{p}}{p!}x_0^p x_{p-1}.\label{eq Vertex diagram free energy}
\end{align}

This may be calculated in old-fashioned perturbation theory as in the previous subsection. The $\alpha$ propagators lead to an effective interaction $V= \frac{\Lambda \pi m^2}{N}\int d^1x(\phi(x)^2)^2$ which may be used to calculate $\hat{T}$ in the DMB formula. For any given diagram with $p$ interaction vertices, the field operators may be contracted in $p!$ different orderings, which involve different energies of intermediate states.

 The most singular contribution is due to the monotonic ordering such that each intermediate state has the same energy as the external particles $E_\alpha$. Each propagator $G_0$ evaluates to $(E-E_\alpha+i\epsilon)^{-1}$. This means that for the purposes of calculating this contribution we can treat $G_0$ like a c-number in the DMB formula,
\begin{align*}
\log \hat{S} =\log\left(	1+	\frac{\left(G_0-\bar{G}_0\right)V}{1-G_0 V}\right) = \log\left(1-\bar{G}_0 V\right)-\log\left(1-{G}_0 V\right).
\end{align*}
Now the logarithms may be expanded up to $p$th order in $V$,
\begin{align}
\left(\log \hat{S}\right)^{(p)}= V^p\frac{1}{p}\left(G_0^p-\hat{G}_0^p\right)=V^p\frac{(-1)^{p-1}}{p!}\frac{d^{p-1}}{dE^{p-1}}\left(G_0-\bar{G}_0\right).\label{eq on shell most singular term}
\end{align}

So the DMB formula for the most singular ordering leads to a $(p-1)$th derivative of an energy delta function. This will correspond to the $l=0$ term in the expansion \eqref{eq xS expansion} for $x_{p-1}$. The other orderings which are less singular will lead to lower-order derivatives of a delta function, and these will correspond to the $l>0$ terms in the expansion. This is shown in detail for $p=4$ in Appendix \ref{Sec Appendix large N alpha4}.

In the remainder of this section we will continue to focus on the most singular ordering, and show agreement with the expression \eqref{eq Vertex diagram free energy} for the free energy $f_{\alpha^p}$. After integrating by parts, the $(p-1)$th derivative leads to a factor of $\beta^{p-1}$, and the DMB formula reduces to
\begin{align}
	f_{\alpha^p}=  \frac{1}{L}\frac{N^{p+1}}{(p+1)!}\int \left(\prod_{\alpha}^{p+1} \frac{dk_\alpha}{(2\pi)2E_\alpha}\right)e^{-\beta \sum_\alpha E_\alpha}\left\langle V^p\right\rangle_{\alpha^p} \frac{(-1)^{p-1}}{p!}\beta^{p-1}.
\end{align}
$\left\langle V^p\right\rangle_{\alpha^p}$ indicates the factors arising from the contraction of $V^p$ as in the forward scattering diagram in Fig \ref{Fig VertexDiagrams}. This factor is the same for all orderings,
\begin{align}
\left\langle V^p\right\rangle_{\alpha^p}=\left(8\times \frac{\Lambda \pi m^2}{N}\right)^p\frac{(p+1)!}{(2E')^{p-1}}L.
\end{align}
There are $p$ vertex factors from the effective interaction $V$, each introducing a combinatorial factor of $8$. There are $(p+1)!$ permutations of the large $N$ indices of the particles, a factor of $L$ from the momentum conservation delta function, and factors of $(2E')^{-1}$ from contracting the fields in the $p-1$ internal propagators.

In total
\begin{align}
		f_{\alpha^p}=-\frac{Nm^2}{\pi}\frac{(-2\Lambda)^p}{p!}\left(\int \,\frac{dk}{2E} e^{-\beta E}\right)^p \int \frac{dk'}{2E'}\frac{m^{2(p-1)}}{\left(E'\right)^{p-1}}\beta^{p-1}e^{-\beta E'}
\end{align}
which indeed agrees with \eqref{eq Vertex diagram free energy} and the $l=0$ term of \eqref{eq xS expansion} at lowest order in fugacity.

	\section{The TBA and bound states}\label{Sec TBA}
	Although our motivation is coming from more general non-integrable theories like the $\phi^4$ theory, it is useful to consider a perturbative expansion of the TBA in order to have an extra check on the validity of the DMB approach. In Sec \ref{Sec TBA LL} we will consider the TBA for the attractive Lieb-Liniger model and show that the perturbative expansion is consistent for both signs of the coupling. The collinear contributions to the free energy cancel delicately with the contributions coming from the bound states that appear for attractive coupling. In Sec \ref{Sec TBA DMB bound} the DMB approach is used to calculate the density of states in the two-particle sector of the attractive LL model, and it is shown how the two-body bound state arises.

	\subsection{Perturbative expansion of the TBA}\label{Sec TBA perturbative}
	We have already compared the DMB formula and the TBA at second order in fugacity back in Sec \ref{Sec DMB diagonal S matrix}. Now to consider the details of three-body forward scattering contributions it is useful to instead expand the TBA to second order in the coupling but to all orders in fugacity. 
	
	The two-body and three-body parts of the free energy density can be distinguished by the number of appearances of the kernel. The two-body part \eqref{TBA free energy two body} has already been suggested from the fugacity expansion,
	\begin{align}
		f_{\text{2}}^{(2)}=+T\int \frac{dp}{2\pi} \frac{1}{e^{\beta E}-1}\left[\log\left(1-e^{-\beta E}\right)\star K^{(2)}\right].\label{TBA free energy two body 2nd}
	\end{align}
The second-order three-body part is
	\begin{align}
	f_3^{(2)}&= -T\int \frac{dp}{2\pi}\frac{1}{e^{\beta E}-1}\left(\frac{1}{2}\frac{\left(\log\left(1-e^{-\beta E}\right)\star K^{(1)}\right)^2}{1-e^{-\beta E}}-\left[\frac{1}{e^{\beta E}-1}{\left(\log\left(1-e^{-\beta E}\right)\star K^{(1)}\right)}\right]\star K^{(1)} \right)\non
		&=+\frac{T}{2}\der{}{\beta}\int \frac{d\theta}{2\pi}\frac{1}{e^{\beta E}-1}\left(\log\left(1-e^{-\beta E}\right)\star K^{(1)}\right)^2.\label{TBA free energy three body}
	\end{align}
	The second, simpler form of $f_{3}$ is valid for SG but not LL.
	\subsubsection{Perturbative expansion of the kernel}
		Some care is required in defining the perturbative expansion of the kernel. Recall that the kernel for both SG and LL is proportional to \begin{align}
		K_{LL} =\frac{1}{\pi}\frac{c}{p^2+c^2}\simeq \frac{c}{\pi p^2}\sum_{j=0}^\infty\left(-\frac{c^2}{p^2}\right)^j.\label{kernel expansion naive}
	\end{align}
	
	This perturbative expansion is valid for $p\neq 0$, but it fails at $p=0$ and so can not be used for calculating the collinear contributions, such as the melon diagram. The divergence of the $K^{(1)}$ term at $p=0$ must be regularized, and there will also be additional terms involving derivatives of delta functions for even powers of $c$.
	
		A valid perturbative expansion of the kernel that takes the behavior near $p=0$ into account may be found by simply Fourier transforming before expanding. This is easy to see for the LL kernel,
	\begin{align*}
		K_{LL}(p)=\int \frac{du}{2\pi}e^{-c|u|}e^{-ipu}=\sum_{j=0}^\infty \frac{\left(-1\right)^jc^j}{j!}\int \frac{du}{2\pi}|u|^je^{-ipu}.
	\end{align*}
	Even powers of $j$ are just even derivatives of delta functions. Odd powers of $j$ involve odd derivatives of the inverse Fourier transform of $\text{sgn}\,u$, which may be understood as the principal value of $1/p$,
	$$\mathcal{P}\frac{1}{p}=\frac{p}{p^2+\epsilon^2}=\pi i\int \frac{du}{2\pi}\text{sgn}\,u\,e^{-\epsilon|u|}e^{-ipu}.$$
	So this gives a definite regularization at $p=0$ for the inverse powers of $p^2$ in the naive expansion of the kernel \eqref{kernel expansion naive}.
	
	The situation is qualitatively the same for the SG kernel.
	\begin{align}
		K_{SG}(\theta)=\frac{1}{2\pi}\frac{2(c/m)\cosh\theta}{\sinh^2\theta+(c/m)^2}=\int \frac{d\nu}{2\pi}\frac{e^{-\gamma|\nu|}+e^{-\left(\pi-\gamma\right)|\nu|}}{1+e^{-\pi|\nu|}}e^{-i\nu\theta},
	\end{align}
	where $\gamma\equiv \arcsin (c/m)$. This may be expanded to $\order{c^2}$,
	\begin{align}
		\frac{e^{-\gamma|\nu|}+e^{-\left(\pi-\gamma\right)|\nu|}}{1+e^{-\pi|\nu|}}=1-c\tanh\frac{\pi\nu}{2}\left(\frac{\nu}{m}\right)+\frac{c^2}{2}\left(\frac{\nu}{m}\right)^2+\order{c^3}.\label{sg kernel fourier transform}
	\end{align}
	Recall that $c= \frac{m g^2}{8}-\frac{m g^4}{64\pi}+\order{g^6}$, 	so the total second-order SG kernel is,
	\begin{align}
		K_{SG}^{(2)}=-\frac{g^2}{8\pi}K_{SG}^{(1)}-\frac{g^4}{128}\delta^{''}(\theta).
	\end{align}
	
	When this is substituted in \eqref{TBA free energy two body 2nd}, the first term in $K^{(2)}_{SG}$ leads to the two-body part of the bubble diagram \eqref{f bubble 2-body} and the second term leads to the two-body part of the melon diagram \eqref{f melon 2-body}.
	
	\subsubsection{The bubble and melon from the TBA}
	
	The three-body contribution $f_3^{(2)}$ involves the first-order kernel
	\begin{align}
		K^{(1)}_{SG}= -\frac{g^2}{8\pi}\derOrd{}{\theta}\mathcal{P} \frac{1}{\sinh \theta}.
	\end{align}
	The principal value regularization appears since the function $\pi i\tanh (\pi\nu/2)$ in \eqref{sg kernel fourier transform} may be understood as the Fourier transform of  $\mathcal{P}\left(\sinh\theta\right)^{-1}$.\footnote{The exact Fourier transform for finite $\epsilon$ is $\frac{\sinh\theta}{\sinh^2\theta +\epsilon^2}=	\pi i \int \frac{e^{-\delta\nu}-e^{-\left(\pi-\delta\right)\nu}}{\left(1+e^{-\pi\nu}\right)\cosh\delta}e^{-i\nu\theta}$ with $\delta\equiv \arcsin \epsilon$.}
	
First let us naively calculate $f_3^{(2)}$ neglecting the details of the principal value regularization.  We calculate \eqref{TBA free energy three body}, using the notation $p_{12}\equiv m\sinh (\theta_1-\theta_2)$,
\begin{align}
	f_{3}^{(2)}&\simeq	\frac{T}{2}\left(\frac{mg^2}{4}\right)^2\der{}{\beta}\int\frac{d\theta_1d\theta_2d\theta_3}{(2\pi)^3}\left(e^{\beta E_1}-1\right)^{-1}\frac{1}{p_{12}p_{13}}\frac{\partial^2}{\partial\theta_2\partial\theta_3}\log\left(1-e^{-\beta E_2}\right)\log\left(1-e^{-\beta E_3}\right)\non
	&\quad =\frac{T}{2}\left(\frac{mg^2}{4}\right)^2\der{}{\beta}\int\frac{d\theta_1d\theta_2d\theta_3}{(2\pi)^3}n_1n_2n_3\frac{p_2 p_3}{p_{12}p_{13}}\beta^2.\nonumber
\end{align}
Now if the momentum factor is symmetrized it can be shown to reduce to $1$,
\begin{align}
	\frac{p_2 p_3}{p_{12}p_{13}}+\frac{p_3 p_1}{p_{23}p_{21}}+\frac{p_1 p_2}{p_{31}p_{32}}=1,\label{magic TBA identity}
\end{align}
\begin{align}
f_{3}^{(2)}&\simeq \frac{T}{3!}\left(\frac{mg^2}{4}\right)^2\der{}{\beta}\int\frac{d\theta_1d\theta_2d\theta_3}{(2\pi)^3}n_1n_2n_3\beta^2.\label{f phi6 and bubble}
\end{align}
The $\beta$ derivative acting on $\beta^2$ produces the free energy due to the sextic vertex diagram \eqref{f phi6 3-body}, and the $\beta$ derivative acting on one of the occupation numbers produces the free energy due to the bubble diagram \eqref{f bubble 3-body}.

	The melon diagram is due to the details of the principal value prescription. Using identities like $\mathcal{P}\frac{1}{p_{12}}=\frac{1}{p_{12}+i\epsilon}+i\pi\delta \left(p_{12}\right)$, all momentum denominators may be adjusted so that they have the same sign of $i\epsilon$. The regularized version of \eqref{magic TBA identity} is
	\begin{align}
		&p_2 p_3\mathcal{P}\frac{1}{p_{12}}\mathcal{P}\frac{1}{p_{13}}+p_3 p_1\mathcal{P}\frac{1}{p_{23}}\mathcal{P}\frac{1}{p_{21}}+p_1 p_2\mathcal{P}\frac{1}{p_{31}}\mathcal{P}\frac{1}{p_{32}}\non&\qquad=\frac{p_{12}p_{13}p_{23}+\order{\epsilon}}{(p_{12}+i\epsilon)(p_{13}+i\epsilon)(p_{23}+i\epsilon)}+2\pi ip_1p_2\delta(p_{13})\mathcal{P}\frac{1}{p_{12}}+\frac{p_1^2}{4}(2\pi)^2\delta(p_{12})\delta(p_{13}).
	\end{align}
	The first term reduces to $1$, and leads to \eqref{f phi6 and bubble} as before. The second term is odd under the transformation $p_1, p_2 \rightarrow -p_1, -p_2$ and thus vanishes upon integration. It is the third term that leads to the three-body part of the melon diagram \eqref{f melon 3-body},
	\begin{align*}
		f_{\text{melon},\,3}&= \frac{T}{3!}\left(\frac{mg^2}{4}\right)^2\der{}{\beta}\int\frac{d\theta}{2\pi}n(\theta)^3 \frac{p^2}{4}\beta^2=-\frac{1}{24}\frac{m^2g^4}{16}\int \frac{d\theta}{2\pi}n(\theta)^3.
	\end{align*}
	
	\subsection{The attractive Lieb-Liniger model}\label{Sec TBA LL}
	The perturbative expansion of the TBA raises a minor puzzle. Upon flipping the sign of the coupling $c$, the kernel $K_{LL}$ flips sign, and thus the two-body correction $f_2$ in \eqref{TBA free energy two body} flips sign. But in the thermal field theory approach, the even orders of perturbation theory do not depend on the sign of $c$, so how is this consistent?
	
	The resolution of course is that the theory with negative $c$ has bound states, and these must explicitly be taken into account in the formulation of the TBA. The contribution from the bound states should compensate the change in sign of the collinear contributions, so that the odd orders of perturbation theory flip sign and the even orders are invariant.
	
	So far we have been focusing mostly on the sinh-Gordon model, and a change in the sign of $c$ corresponds to considering the sine-Gordon model instead. The TBA for the sine-Gordon model \cite{sinGordonGenHydro2024}  is complicated by the presence of the soliton sector. In the following sections we will focus on the simpler LL model instead (see Appendix \ref{Sec Appendix ll} for the LL analogue of calculations in Sec \ref{Sec Intro diagrams bubble melon})
	
The LL model with attractive coupling is not stable since there are bound states with arbitrarily large negative energy. However we may still make sense of the model by truncating at some finite order in the fugacity expansion. The TBA may be constructed using standard methods (see e.g. \cite{tbaIntroTongeren}). There is a distinct pseudoenergy $\epsilon_j$ for each $j$-body bound state,
	\begin{gather}
	f=-T\sum_j j \int \frac{dp}{2\pi} \log\left(1+e^{-\beta\epsilon_j}\right)\label{TBA LL attractive free energy}\\
	\epsilon_j = E_j-T\sum_k \log \left(1+e^{-\beta\epsilon_k}\right){\star}K_{kj},\qquad K_{kj}\equiv \frac{1}{2\pi i}\left.\derOrd{}{u}\log S_{kj}(u)\right|_{u=p_k - p_j}.\label{TBA LL attractive}
\end{gather}
Here $E_j$ is the energy of the $j$-body bound state, and $S_{kj}$ is the two-body S-matrix between a $k$-body bound state and a $j$-body bound state. Only the lowest order quantities will be needed for our purposes
\begin{gather}
	E_1=\frac{p^2}{2m},\quad	E_2=\frac{2p^2}{2m}-\frac{c^2}{4m},\quad E_3=\frac{3p^2}{2m}-\frac{c^2}{m},\\
S_{11}(u)=\frac{u+i|c|}{u-i|c|}, \quad S_{21}(u)=S_{11}(u-{ic}/2)S_{11}(u+{ic}/{2}).
\end{gather}
By expanding \eqref{TBA LL attractive}, the free energy at second-order in fugacity is
\begin{align}
	f_{(2)}=-T\int \frac{dp}{2\pi}\left[-\frac{1}{2}e^{-2\beta E_1}+2e^{-\beta E_2}+e^{-\beta E_1}\left(e^{-\beta E_1}\star K_{11}\right)\right].\label{TBA attractive f2}
\end{align}
In the following section this will be derived in the DMB approach, but first let us consider the perturbative expansion in $c$.

At zeroth order in $c$, this ought to be given by the free boson gas \eqref{free boson gas},
\begin{align*}
	f_{(2)}^{(0)}=-T\int \frac{dp}{2\pi}\frac{1}{2}e^{-2\beta E_1}.
\end{align*}
Indeed $E_2^{(0)}=2E_1$ and $K_{11}^{(0)}(p) =-\delta(p)$, so this works out. Note that even though the theory is not interacting at zeroth order, the bound state contribution is needed to get the appropriate statistics of the free boson gas. The contributions to the free energy \eqref{TBA LL attractive free energy} due to $\epsilon_1$ and $\epsilon_2$ alone do not have the appropriate statistics.

At second order in $c$, the free energy is given by the non-relativistic analogue \eqref{f melon LL} of the two-body part of the melon diagram \eqref{f melon 2-body}
\begin{align*}
	f_{(2)}^{(2)}=-T\int \frac{dp}{2\pi}\frac{\beta c^2}{4m}e^{-2\beta E_1}.
\end{align*}
For positive $c$ this is entirely due to the contribution of the kernel $K^{(2)}$, but for negative $c$ this flips sign
\begin{align*}
-T\int \frac{dp}{2\pi}e^{-\beta E_1}\left(e^{-\beta E_1}\star K_{11}^{(2)}\right)=+T\int \frac{dp}{2\pi}\frac{\beta c^2}{4m}e^{-2\beta E_1}.
\end{align*}
In order to get the appropriate free energy it must cancel with the binding energy of the two-body bound state $\left(e^{-\beta E_2}\right)^{(2)}= \frac{\beta c^2}{4m}e^{-2\beta E_1}$. In this sense, the collinear contributions to the free energy have an intimate relation to the contributions due to bound states for negative $c$.

At higher orders in fugacity similar cancellations occur with the binding energy for higher order bound states. Introducing notation $z_j \equiv e^{-\beta E_j}$, the next-order contribution is
\begin{align}
	f_{(3)}=-T\int \frac{dp}{2\pi}&\left[z_1\left(\frac{1}{3}z_1^2+\left(z_1\left(z_1\star K_{11}\right)\right)\star K_{11}+\frac{1}{2}\left(z_1\star K_{11}\right)^2\right)\right.\non
	&\quad + z_1\left(-z_1\left(z_1\star K_{11}\right)-\frac{1}{2}\left(z_1^2\star K_{11}\right)\right)\non
	&\quad\left. +3 z_3+2z_2\left(z_1\star K_{21}\right)+z_1\left(z_2\star K_{21}\right)\right].
\end{align}
The first line does not depend on the sign of $c$, and the second line flips sign. The third line involves the two-body and three-body bound states, and it is only present for negative $c$. Using the expansion of the kernel $K_{21}$,\begin{align}
	K_{21}(p)=-2\delta(p)+2K_{11}^{(1)}(p)+\frac{5}{4}c^2\delta^{\prime\prime}(p)+\order{c^3},
\end{align}
it can be shown that the perturbative expansion of $f_{(3)}$ indeed has the appropriate dependence on the sign of $c$ at each order.
	\subsection{Bound states in the DMB approach}\label{Sec TBA DMB bound}
	Now we will treat the two-particle sector of the attractive LL model from the DMB approach. This is arguably the simplest example involving bound states. The Hamiltonian is just
	\begin{align}
		H =\frac{p^2}{m}+\frac{c}{m}\delta(x)+\frac{P^2}{4m},
	\end{align}
where the the relative momentum $p=(p_1-p_2)/2$ and position $x=x_1-x_2$ form a canonically conjugate pair, and the total momentum $P=p_1+p_2$ decouples. For the moment, we will ignore the decoupled $P$ degree of freedom.
	
	Using the notation $|x_0\rangle$ for the state localized at $x=0$, the delta function operator in the interaction may be written as $(c/m)|x_0\rangle\langle x_0|$. The T-matrix operator \eqref{Def T} may be explicitly calculated
	\begin{align}
		\hat{T}=\frac{c}{m-\langle x_0| G_0 |x_0\rangle}|x_0\rangle\langle x_0|,
	\end{align}
and the trace-log of the S-matrix becomes
\begin{align}
	\text{Tr}\log \hat{S}=\log\left(1+c \frac{\langle x_0| G_0-\bar{G}_0|x_0 \rangle}{m-\langle x_0|G_0|x_0\rangle}\right).\label{eq tr log S prelim}
\end{align}
For these calculations, note that  $\langle x_0|p\rangle=1$, and thus $\int \frac{dp}{2\pi}\langle p| G_0|x_0 \rangle=\langle x_0| G_0|x_0 \rangle $.

The expectation of the propagator is
\begin{align}
	\langle x_0| G_0|x_0 \rangle = \int \frac{dp}{2\pi} \frac{1}{E-\frac{p^2}{m}+i\epsilon} = -\frac{im}{2\sqrt{mE^+}},
\end{align}
where $\sqrt{mE^+}$ is defined to be the square root of $mE+i\epsilon$ with positive imaginary part. Similarly, $\langle x_0| \bar{G}_0|x_0 \rangle= im /(2\sqrt{mE^-})$, where $\sqrt{mE^-}$ is defined as the square root of $mE-i\epsilon$ with negative imaginary part.

So \eqref{eq tr log S prelim} becomes,
\begin{align}
	\text{Tr}\log \hat{S}=\log\left(1-\frac{ic/2}{\sqrt{mE^+}+ic/2}\left(1+\frac{\sqrt{E^+}}{\sqrt{E^-}}\right)\right).\label{eq tr log S}
\end{align}
In the limit of small $\epsilon$, the factor $\left(1+\frac{\sqrt{E^+}}{\sqrt{E^-}}\right)$ evaluates to $2$ for $E>0$, and vanishes for $E<0$. However if $c$ is negative, then there is a pole at $\sqrt{m E^+}=-ic/2$, and this can compensate the vanishing of the factor. The argument of the logarithm winds around the complex plane in the vicinity of the pole, and this will lead to a delta function localized at the bound state energy after taking the $E$ derivative in \eqref{Eq DMB formula},
$$\frac{1}{2\pi i}\derOrd{}{E}\text{Tr}\log \hat{S}= \frac{dp}{dE}\frac{1}{2\pi}\frac{c}{p^2+c^2/4}\theta(E)+\delta\left(E+\frac{c^2}{4m}\right).$$

Let us now reinstate the total momentum $P$ degree of freedom in the problem. All appearances of $E$ become $E-P^2/4m$, and there is an overall factor of $L$ and an additional $P$ integral due to the trace,
\begin{align}
	\delta Z = L\int dP\left[e^{-\beta\left(\frac{P^2}{4m}-\frac{c^2}{4m}\right)}+\int dp\, e^{-\beta\left(\frac{p^2}{m}+\frac{P^2}{4m}\right)}\frac{1}{2\pi}\frac{2c}{4p^2+c^2} \right].
\end{align}
After changing variables to $\bar{p}\equiv P/2$ in the bound state contribution, to $p_1, p_2$ in the scattering contribution, and multiplying by $-(\beta L)^{-1}$ to obtain the free energy density, this takes the form of $\delta f_{(2)}$ calculated from the TBA in \ref{TBA attractive f2}.
	\section{Discussion}\label{Sec Discussion}
	
	In this paper we have investigated the DMB formula \eqref{Eq DMB formula} for some massive scalar field theories in two spacetime dimensions. In particular we focused on integrable theories for which we may compare with the TBA or a large $N$ expansion. For multi-particle scattering it was shown that there are two issues that complicate a naive treatment of the DMB formula. The first issue has to do with the presence of forward scattering divergences. This issue has been discussed from the outset of the DMB formula \cite{dmb, dm1970, dm1971, norton1986elementary}, but we show here how by simply applying the DMB formula consistently the forward scattering divergences can be associated to derivatives of energy delta functions, and how the sum over all vertex time orderings corresponds to an easily calculable loop diagram in the thermal field theory approach.
	
	The second issue is perhaps specific to two spacetime dimensions and has not to my knowledge been previously discussed in the literature. The issue has to do with the fact that the trace in the DMB formula has large contributions from states with collinear momenta. This issue arises already in a calculation of the melon diagram \eqref{Eq melon full} at second order in the $\phi^4$ theory. It is not just a difficulty of the DMB approach, and calculations of the melon diagram using thermal field theory or the TBA also involve subtleties at collinear momentum. The lesson is that regions of integration with collinear momenta should also be considered carefully for higher order diagrams, and for more complicated 1+1 dimensional theories.
	
	There have not been many concrete examples of application of the DMB approach, so there are a number of possible directions to build on the initial investigations in this paper. With an eye towards the effective string theory program, clearly it would be illuminating to consider scattering in massless 1+1 dimensional theories. Shortly after the publication of the first version of this paper on arXiv, the related paper \cite{BaratellaEliasMiroGendy2024} took the first steps towards this by applying the DMB approach to the integrable $D=3$ effective string theory. This is an important step forward, and ironically the massless $D=3$ theory avoids some of the subtleties seen in the massive theories considered here. But there is still work to be done since the effective string theory for Yang-Mills is not integrable \cite{cooperEtAlConfiningStringIntegrability}\cite{chenEtAlFluxTubesTTbar2018}.
	
	One advantage of the TBA for the effective string theory program is that higher energy levels in finite volume may be calculated \cite{doreyTateo1996}, not just the ground state. It would be interesting to consider higher energy levels in the DMB approach as well. Something similar has already been done for the thermal self-energy \cite{jeonEllis1998}, which gives information on the first excited state in finite volume. Indeed in section \ref{Sec On shell} the energy of first excited state $M(\beta)$ was shown to be given at lowest order by the two-body forward scattering amplitude, as in the work of L\"{u}scher \cite{luscher1984,luscher1986 I}, and it is straightforward to extend this to multi-particle scattering by calculating higher order corrections to the thermal self-energy using the rules of Sec \ref{Sec On shell diagrams}.
	
	We have mostly considered theories with one type of particle in the spectrum which corresponds neatly to the field in the Lagrangian, but of course the possibilities in quantum field theory are much richer. The particles of the free theory may form bound states or may decay into other stable particles when an interaction is included. We have taken a preliminary look at bound states in the Lieb-Liniger model and seen that they may indeed be calculated in the DMB approach. The analysis of Sec \ref{Sec TBA} could be further extended to treating the breather bound states in the sine-Gordon model, although that model is more complicated due to the presence of solitons, and relatedly the presence of non-trivial semi-classical saddles in the thermal field theory approach.
	
	Of course we may avoid these problems for sine-Gordon by considering the dual massive Thirring model instead. Treating fermions in the DMB approach is no major difficulty (see e.g. \cite{norton1986elementary}), and although the calculation was not included here, indeed the lowest order correction to the free energy may be derived from the massive Thirring two-body forward scattering amplitudes. Even so, it may be enlightening to consider the bosonic description of the theory and investigate how the non-perturbative corrections due to the solitons can possibly arise in the DMB framework.
	
	\section*{Acknowledgements}
	I would like to thank Vladimir Rosenhaus for introducing reference \cite{dmb}, and collaboration on an early version of this project. I would also like to thank Bal\'{a}zs Pozsgay for discussions. 
	This work  is supported in part by NSF grant PHY-2209116  and by the ITS through a Simons grant. 
	\appendix
		\section{More on the melon diagram}\label{Sec Appendix melon}
		This section deals more with the melon diagram integral
		$M_4\equiv \int d\tau dx D(\tau,x)^4,$
		where $D(\tau,x)$ is the finite temperature propagator (with $0\leq \tau <\beta$)
			\begin{align}
		D(\tau,x)&=\frac{1}{\beta}\sum_n\int \frac{dk}{2\pi}\frac{e^{-i\left(kx+\omega_n \tau\right)}}{\omega_n^2+k^2+m^2}\non
		&=\int \frac{dk}{2\pi}e^{-ikx}\left(\frac{e^{- E \left|\tau\right|}+e^{- E\left(\beta- \left|\tau\right|\right)}}{2E\left(1-e^{-\beta E}\right)}\right). \label{Def D}
	\end{align}
	
	For a theory with interaction term $\frac{m^2g^2}{4!}\phi^4$, the contribution to the free energy density is
	\begin{align}
		f_{\text{melon}}=-\frac{1}{3}\frac{m^4g^4}{16}M_4-\text{counterterms},
	\end{align}
	where the counterterms subtract the zero-body and one-body parts.
	
	The full result for $M_4$ is
		\begin{align}
			M_4 = \frac{7}{8m^2}\frac{\zeta(3)}{(2\pi)^3}+\frac{1}{16m^2}\int\frac{d\theta}{2\pi}\left[n(\theta)+3n(\theta)^2+2n(\theta)^3\right].\label{Eq melon full}
		\end{align}
	The two-body part was calculated in \eqref{f melon 2-body}. The three-body part \eqref{f melon 3-body} was calculated indirectly via the TBA in Sec \ref{Sec TBA perturbative}, and will be calculated directly here. 
		
		\subsection{Zero- and one-body parts}
		For completeness we will also present the parts of the integral $M_4$ that are canceled by counterterms.
		
		At zero temperature ($\beta\rightarrow\infty$), the propagator is just given by a Bessel function,
		\begin{align}
			D_\infty(\tau,x)=\frac{1}{2\pi}K_0(m\sqrt{\tau^2+x^2}).
		\end{align}
		The `zero-body' part is equivalent to the zero-temperature limit,
		\begin{align}
			M_{4,\,0}=\frac{1}{m^2(2\pi)^3}\int_0^\infty du\,u\,K_0(u)^4 =\frac{7}{8m^2}\frac{\zeta(3)}{(2\pi)^3}.
		\end{align}
		
		The one-body part involves the self-energy diagram given by cutting one line of the melon diagram. This self-energy diagram is often referred to as the `sunset' diagram and it is given by
		\begin{align}
			\Pi_{\text{sunset}}(p)&=
			\int d^2x e^{ip\cdot x}D_\infty(x)^3 = \frac{1}{\left(2\pi\right)^2}\int_0^\infty dr \,r J_0\left(|p|r\right)\left(K_0(mr)\right)^3.
		\end{align}
		The one-body part of $M_4$ is then given by the on-shell ($p^2=-m^2$) value of the sunset diagram,
		\begin{align}
			M_{4,\,1}&=\int\frac{d\theta}{2\pi}n(\theta)\left.4\Pi_{\text{sunset}}\right|_{p^2=-m^2}\non&=\int\frac{d\theta}{2\pi}n(\theta)\left(\frac{4}{m^2\left(2\pi\right)^2}\int_0^\infty du \,u\, I_0\left(u\right)\left(K_0(u)\right)^3\right)\non
			&=\frac{1}{16m^2}\int\frac{d\theta}{2\pi}n(\theta).
		\end{align}
		\subsection{Cutting the melon diagram}\label{Sec Appendix melon cut}
		
		For the sake of finding the two- and three-body parts, cutting the melon diagram according to the method of Sec \ref{Sec Intro diagrams general cutting} is not very convenient due to the singularities in $\mathcal{M}_\pi$. In Sec \ref{Sec Intro diagrams bubble melon} we nevertheless considered the forward scattering amplitudes corresponding to cutting a line of the melon diagram, and regularized them by the principal value prescription suggested by the second-order DMB formula \eqref{DMB principal value}.
		
		In this subsection we will justify this directly from the integral $M_4$. This involves integrating the vertices of the diagram over Euclidean time, rather than summing over Matsubara frequencies. There is a generalization of this method \cite{norton1986elementary,jeon1993} based on the work of Baym and Sessler \cite{baymSessler1963}, but $M_4$ is particularly simple, and it has been considered in \cite{bugrijShadura1995,massiveBasketball}.
		
		The propagator \eqref{Def D} may be written
		\begin{align}
			D(\tau,x)=\int \frac{dk}{2\pi}e^{ik\cdot x}\int \frac{d\omega}{2E}\left(\delta(\omega-E)-\delta(\omega+E)\right)\left(\frac{e^{- \omega \left|\tau\right|}}{1-e^{-\beta \omega}}\right),\label{Eq D form 2}
		\end{align}
	and a melon diagram $M_N$ with $N$ propagators going between two vertices  may be calculated,
		\begin{gather}
			M_N\equiv\int d\tau dx D(\tau,x)^N=\int \frac{dk_1}{2\pi}\dots \frac{dk_N}{2\pi}2\pi \delta\left(\sum_i k_i\right) \tilde{M}_N,\\
			\tilde{M}_N=\int \prod_{j=1}^N\left( \frac{d\omega_j}{2E_j}\frac{\delta(\omega_j-E_j)-\delta(\omega_j+E_j)}{1-e^{-\beta \omega_j}}\right)\frac{1-e^{-\beta\sum_i\omega_i}}{\sum_i\omega_i}.
		\end{gather} 
		
		Let us focus on $\tilde{M}_4$,
		\begin{align}
			\tilde{M}_4&=2\left(\prod_{j=1}^4\frac{n_j}{2E_j}\right)\left[\frac{e^{\beta\sum_iE_i}-1}{\sum_iE_i}+4\frac{e^{\beta \left(E_1+E_2+E_3\right)}-e^{\beta E_4}}{E_1+E_2+E_3-E_4}+3\frac{e^{\beta \left(E_1+E_2\right)}-e^{\beta\left(E_3+E_4\right)}}{E_1+E_2-E_3-E_4}\right].\label{M4tilde}
		\end{align}
	The final term of $\tilde{M}_4$ is actually analytic, since the numerator vanishes when $E_1+E_2=E_3+E_4$. Thus it is valid to adjust the denominator by an arbitrary infinitessimal $\pm i\epsilon$, after which we may exchange integration variables $1,2$ and $3,4$,
	\begin{gather*}
		3\frac{e^{\beta \left(E_1+E_2\right)}-e^{\beta\left(E_3+E_4\right)}}{E_1+E_2-E_3-E_4\pm i\epsilon}\rightarrow\mathcal{P}\frac{6\,e^{\beta \left(E_1+E_2\right)}}{E_1+E_2-E_3-E_4}.\label{M4tildeP}
	\end{gather*} 
	This is ultimately how the principal value prescription \eqref{DMB principal value} arises in the thermal field theory approach.
	
		The exponentials in the numerators may be written as $e^{\beta E_j}= n_j^{-1}+1$ and so some of the occupation numbers in the product $\prod_j n_j$ will be canceled. The $r$-body part of the diagram is just taken to be the part of the diagram with $r$ overall occupation number factors. The three-body part is
		\begin{align}
			\tilde{M}_{4,\,3}&=\left(\prod_{j=1}^3\frac{n_j}{E_j}\right)\left[\frac{1}{-E_4^2+\left(E_1+E_2+E_3\right)^2}+\mathcal{P}\frac{3}{-E_4^2+\left(E_1+E_2-E_3\right)^2}\right].\label{M4tilde 3-body}
		\end{align}
		This obviously agrees with the DMB expression \eqref{f melon 3-body prop}.
		
		The two-body part is
		\begin{align}
			\tilde{M}_{4,\,2}&=\frac{3}{4}\frac{n_1 n_2}{E_1 E_2 E_3 E_4}\left[\frac{1}{\sum_iE_i}+\frac{2}{E_3+E_4+E_1-E_2}+\mathcal{P}\frac{1}{E_3+E_4-E_1-E_2}\right].\label{M4tilde 2-body}
		\end{align}
		This also agrees with the DMB expression \eqref{f melon 2-body J} after we write it in terms of the integral $J(p)$, which may be written in terms of energy denominators
		\begin{align}
			J(p)=\int \frac{dk_3}{2\pi}\frac{dk_4}{2\pi}\frac{2\pi\delta(k_3+k_4-p^1)}{2 E_3\, 2E_4}\left[\frac{1}{ip^0+E_3+E_4}+\frac{1}{-ip^0+E_3+E_4}\right].
		\end{align}
	
		\subsection{Direct calculation of the three-body part}\label{Sec Appendix melon 3body}
		This section builds on the discussion of Sec \ref{Sec Intro diagrams melon 3body}, but the $i\epsilon$ terms are considered carefully.
		
		For brevity introduce notation such as \begin{align}
			e_{12}\equiv e^{\theta_1}-e^{\theta_2},\qquad \deltabar_{12}\equiv 2\pi \delta(e_{12}),\qquad  H_{12}\equiv \frac{\text{sgn}(e_{12})+1}{2}.
		\end{align}
		As usual the Heaviside function $H_{12}$ is taken to equal $1/2$ when $\theta_1=\theta_2$.
		
		From \eqref{naive melon three-body}, the principal value of the scattering amplitude corresponding to the middle diagram on the lower row of Fig \ref{Fig bubble melon} is
		\begin{align}
			\mathcal{P}\frac{e^{\theta_1+\theta_2+\theta_3}}{m^2 \left(e^{\theta_1}+e^{\theta_2}\right)e_{13}e_{23}}&=\frac{e^{\theta_1+\theta_2+\theta_3}}{m^2\left(e^{\theta_1}+e^{\theta_2}\right)}\text{Re}\frac{1}{\left(e_{13}+i\text{sgn}(e_{23})\epsilon\right)\left(e_{23}+i\text{sgn}(e_{13})\epsilon\right)}\non
			&=\frac{e^{\theta_1+\theta_2+\theta_3}}{m^2\left(e^{\theta_1}+e^{\theta_2}\right)}\text{Re}\left[\left(\frac{1}{e_{13}+i\epsilon}+i H_{32}\deltabar_{13}\right)\left(\frac{1}{e_{23}+i\epsilon}+i H_{31}\deltabar_{2 3}\right)\right]\non
			&=\frac{e^{\theta_1+\theta_2+\theta_3}}{m^2\left(e^{\theta_1}+e^{\theta_2}\right)}\left[\left(\frac{1}{e_{13}+i\epsilon}\right)\left(\frac{1}{e_{23}+i\epsilon}\right)+\frac{1}{4}\deltabar_{13}\deltabar_{2 3}+\frac{i}{2}\left(\deltabar_{13}\mathcal{P}\frac{1}{e_{23}}+\deltabar_{23}\mathcal{P}\frac{1}{e_{13}}\right)\right].\label{App 3-body intermediate step}
		\end{align}
		The last terms involving a single delta function will vanish upon integration since for instance,
		\begin{align*}
			\frac{e^{\theta_1+\theta_2+\theta_3}}{e^{\theta_1}+e^{\theta_2}}\deltabar_{13}\mathcal{P}\frac{1}{e_{23}}=\frac{e^{\theta_1+\theta_2}}{e^{\theta_1}+e^{\theta_2}}2\pi \delta(\theta_1-\theta_3)\mathcal{P}\frac{e^{-\frac{\theta_1+\theta_2}{2}}}{2\sinh\frac{\theta_1-\theta_2}{2}}=2\pi \delta(\theta_1-\theta_3)\mathcal{P}\frac{1}{2\sinh{\left(\theta_1-\theta_2\right)}},
		\end{align*}
		and this is an odd under the interchange of the signs of $\theta_1, \theta_2$.
		
		The first term in \eqref{App 3-body intermediate step} will be symmetrized in order to cancel in the same way as in \eqref{naive melon three-body}, but since the signs of the $i\epsilon$ terms must be adjusted to be the same this will also produce some additional terms,
		\begin{align}
			&\frac{1}{3}\left[\frac{1}{e^{\theta_1}+e^{\theta_2}}\left(\frac{1}{e_{13}+i\epsilon}\right)\left(\frac{1}{e_{23}+i\epsilon}\right)+\left(3\leftrightarrow 1\right) + \left(3\leftrightarrow 2\right)\right]\non&\qquad=\frac{1}{3}\left[\dots-\frac{1}{e^{\theta_3}+e^{\theta_2}}i\deltabar_{13}\left(\frac{1}{e_{21}+i\epsilon}\right)+\frac{1}{e^{\theta_1}+e^{\theta_3}}\left(i\deltabar_{12}\left(\frac{1}{e_{23}+i\epsilon}\right) +i\deltabar_{23}\left(\frac{1}{e_{21}+i\epsilon}\right)-\deltabar_{12}\deltabar_{23} \right)\right]\non
			&\qquad=\frac{1}{3}\left[\dots-\frac{1}{2}\frac{1}{e^{\theta_3}+e^{\theta_2}}\deltabar_{13}\deltabar_{12}\right],
		\end{align}
		where the ellipsis includes both the terms which will cancel as in \eqref{naive melon three-body} and the terms with a single delta function that were shown to vanish immediately above.
		
		So in total the non-vanishing part of $M_{4,\,3}$ in \eqref{M4tilde 3-body} involves a double delta function,
		\begin{align}
			M_{4,\,\text{3}} = \int\frac{d\theta_1 d\theta_2 d\theta_3}{(2\pi)^3}n_1^3\frac{3e^{3\theta_1}}{2m^2e^{\theta_1}}\left(\frac{1}{4}-\frac{1}{6}\right)\deltabar_{13}\deltabar_{2 3}=\frac{2}{16m^2}\int\frac{d\theta }{2\pi}n(\theta)^3.
		\end{align}
	
	\section{More on large $N$ diagrams}\label{Sec Appendix large N}

	\subsection{An expansion of $x_s$}\label{Sec Appendix large N xS}
Recall that $x_s$ is defined as a sum of Bessel functions,
	\begin{align}
		x_s\equiv \sum_{j=1}^\infty\left(j\zeta\right)^sK_s(j\zeta),\qquad \zeta\equiv \beta m.
	\end{align}
	There is a formula for $x_s$ in terms of energy derivatives of $n$ which was used in Sec \ref{Sec On shell} to relate thermal field theory diagrams to the DMB formula,
	\begin{align}
		x_s = \int \frac{dk}{2E}\frac{m^{2s}}{E^s}\sum_{l=0}^s \frac{(s+l)!}{(2E)^l(s-l)!l!}\left(-\frac{d}{dE}\right)^{s-l}n(E).\label{eq xS expansion appendix}
	\end{align}

This formula may be proven inductively. The $s=0$ case involves a well-known integral expression for Bessel functions
\begin{align}
	x_0=\sum_{j=1}^\infty K_0(j\zeta)=\sum_{j=1}^\infty \frac{1}{2}\int d\theta \, e^{-j\zeta \cosh \theta}=\int \frac{dk}{2E} \, n(E).
\end{align}

To prove \eqref{eq xS expansion appendix} for $x_{s+1}$, first note that\footnote{Incidentally, the $s=0$ case of this formula can be used to make the simplification \eqref{f bubble 3-body}.}
\begin{align}
	-\zeta^{2s+1}\derOrd{}{\zeta}\zeta^{-2s}x_s=\sum_{j=1}^\infty \left(j\zeta\right)^{s+1}\left(\frac{s}{j\zeta}-\derOrd{}{\left(j\zeta\right)}\right)K_{s}(j\zeta) = x_{s+1}.
\end{align}
The same operator may be written as a derivative of $m^2$ at fixed $\beta$,
$$-\zeta^{2s+1}\derOrd{}{\zeta}\zeta^{-2s}=-2 m^{2s+2}\derOrd{}{m^2}m^{-2s}.$$

Now act with this operator on both sides of \eqref{eq xS expansion appendix},
\begin{align}
	x_{s+1}&=-m^{2s+2}\int dk\derOrd{}{m^2}\left[\frac{1}{E^{s+1}}\sum_{l=0}^s \frac{(s+l)!}{(2E)^l(s-l)!l!}\left(-\frac{d}{dE}\right)^{s-l}n(E)\right].
\end{align}
Inside the integrand, the operator $\left.\derOrd{}{m^2}\right.=\frac{1}{2E}\derOrd{}{E}$ for fixed $k$. It is now a simple exercise to show that the acting with $d/dE$ reproduces the terms of \eqref{eq xS expansion appendix} for $x_{s+1}$, completing the proof.

Also note that the summation over $j$ involved in $x_s$ is not essential to the proof. If $n(E)$ is expanded like $\sum_{j=1} e^{-j\beta E}$, then the formula holds order by order in $j$. In particular it is true at the leading order in fugacity, $j=1$.

\subsection{Calculation of $f_{\alpha^4}$}\label{Sec Appendix large N alpha4}

Recall from Sec \ref{Sec On shell} that the free energy $f_{\alpha^p}$ due to $p+1$ particles scattering in the manner of Fig \ref{Fig VertexDiagrams} is given by \eqref{eq Vertex diagram free energy},
\begin{align*}
	f_{\alpha^p}=-\frac{Nm^2}{\pi}\frac{(-2\Lambda)^{p}}{p!}x_0^p x_{p-1}.
\end{align*}

The leading order ($l=0$) term of this formula was matched to the DMB formula for arbitrary $p$ in Sec \ref{Sec On shell DMB}. In this section we will fix $p=4$, but match the full $l$ expansion in \eqref{eq xS expansion appendix} to the DMB formula. We will continue to work at lowest order in fugacity, corresponding to the leading order $(j=1)$ terms in all appearances of $x_s$.

$f_{\alpha^4}$ involves the factor $x_3$, which is given by \eqref{eq xS expansion appendix},
\begin{align}
	x_{3\,(1)}=	m^6 \int \frac{dk'}{2E'}\frac{e^{-\beta E'}}{{E'}^3}\left(\beta^3+\frac{6\beta^2}{E'}+\frac{15\beta}{{E'}^2}+\frac{15}{{E'}^3}\right).\label{eq x3 expansion}
\end{align}
Since the overall factors in $f_{\alpha^4}$ have already been established for the leading order $\beta^3$ term in Sec \ref{Sec On shell DMB}, it is sufficient here to reproduce the expansion in parentheses.

We can expand $\log \hat{S}$ in the DMB formula up to fourth order. Each term will take the schematic form 
$$\Delta G\, V O_{3} V O_{2} V O_{1} V $$
where $\Delta G\equiv G_0-\bar{G}_0$, and the operator $O_{i}$ at position $i$ represents either a propagator  $G_0$ or a delta function $\Delta G$. When the trace is taken, the $H_0$ in leftmost copy of $\Delta G$ will evaluate to the energy of the incoming or outgoing state $E_\alpha$, but the three operators $O_1, O_2, O_3$ may involve some other intermediate energy which will depend on the `time ordering' in which the four copies of $V$ are contracted (see e.g. Fig \ref{Fig OFPT} for the $p=2$ case).

There are $4!=24$ total time orderings of the vertices of the forward scattering diagram corresponding to $f_{\alpha^4}$.

\begin{itemize}
\item Of these, only the single ordering considered in Sec \ref{Sec On shell DMB higher} will involve all three intermediate energies evaluating to $E_\alpha$. According to \eqref{eq on shell most singular term} this leads to a term proportional to
	$$-\frac{1}{4!}\frac{d^3}{dE^3}\delta\left(E-E_\alpha\right).$$
	\item There are 3 time orderings where two intermediate energies are on-shell, and the remaining energy is $E_\alpha+2E'$, where $E'$ is the energy of the privileged particle in Fig \ref{Fig VertexDiagrams}. Any delta functions $\Delta G$ vanish if they are evaluated at an `off-shell' energy (i.e. an energy not equal to $E_\alpha$), and the terms in the fourth-order expansion of $\log \hat{S}$ that survive are just proportional to a \emph{third-order} expansion of $\log \hat{S}$ with all intermediate energies set to $E_\alpha$. So using \eqref{eq on shell most singular term} again, these 3 time orderings lead to a term
	$$+3\left(\frac{1}{E-E_\alpha-2E'}\right)\frac{1}{3!}\frac{d^2}{dE^2}\delta\left(E-E_\alpha\right).$$
	\item Proceeding similarly, there are 7 time orderings with one energy equal to $E_\alpha$ and two equal to $E_\alpha + 2E'$,
	$$-7\left(\frac{1}{E-E_\alpha-2E'}\right)^2\frac{1}{2!}\frac{d}{dE}\delta\left(E-E_\alpha\right).$$
	\item There are 9 time orderings where all energies equal $E_\alpha + 2E'$, and 4 time orderings where two energies equal $E_\alpha + 2E'$, and one energy equals $E_\alpha + 4E'$,
	$$+\left[9\left(\frac{1}{-2E'}\right)^3+4\left(\frac{1}{-2E'}\right)^2\left(\frac{1}{-4E'}\right)\right]\delta\left(E-E_\alpha\right).$$
\end{itemize}

All of these terms are added together in the density of states, where they multiply $e^{-\beta E}$ in the integrand of \eqref{Eq DMB formula}. After integrating the derivatives of delta functions by parts (being careful to also act on the appearance of $E$ in the off-shell propagator), the sum of the above terms becomes
$$-\frac{1}{4!}\left(\beta^3+\frac{6\beta^2}{E'}+\frac{15\beta}{\left(E'\right)^2}+\frac{15}{\left(E'\right)^3}\right)\delta\left(E-E_\alpha\right).$$
which gives the factor in parentheses in \eqref{eq x3 expansion}, which we set out to show.


	
	\section{The Lieb-Liniger model}\label{Sec Appendix ll}
	\begin{figure}
		\centering
		\includegraphics[width=0.5\textwidth]{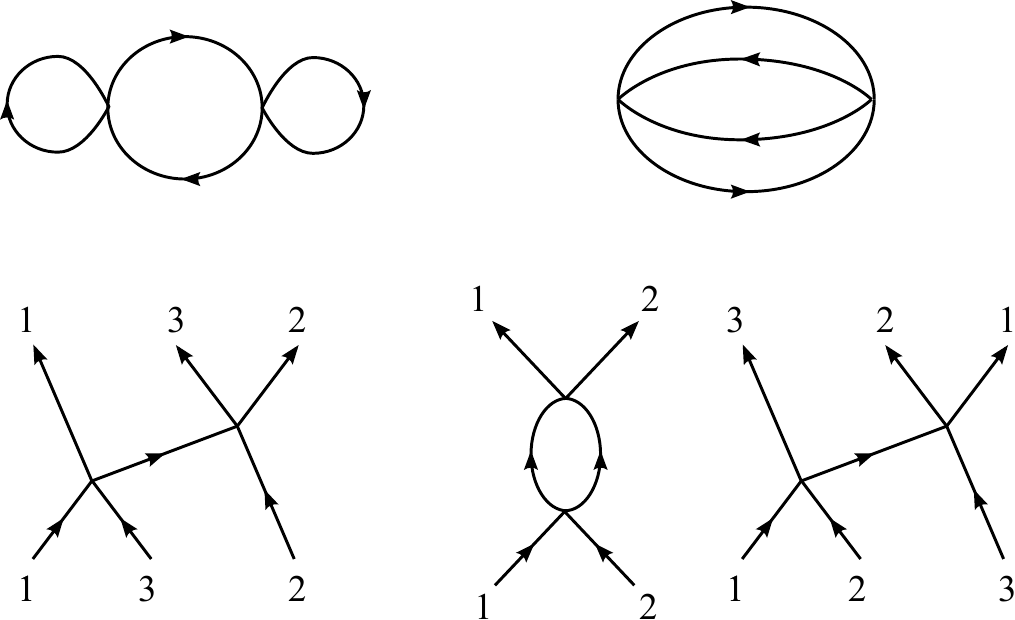}
		\caption{The `bubble' and `melon' corrections to the free energy for the Lieb-Liniger model.}\label{Fig LL}
	\end{figure}
	In the text the discussion has mostly focused on diagrams in the relativistic $\phi^4$ and sinh-Gordon models, but much the same story occurs for the non-relativistic Lieb-Liniger (LL) model, where the two-body interaction involves a repulsive delta function potential of strength $c/m$.
	
	We may calculate with a coherent state path integral with Lagrangian
	\begin{align}
		\Lagr_{LL}=\sum_ka^\dagger_k \left( \partial_\tau + E_k\right) a_k +\frac{c}{2m}\sum_{k+l=m+n}a^\dagger_m a^\dagger_n a_k a_l,
	\end{align}
	where the indices $k$ on the c-number field $a_k$ refer to spatial momentum. No counterterms are needed as long as the $a^\dagger$ fields in the interaction vertex are understood to be at an infinitesimally later time than the $a$ fields. The propagators $\langle a(\tau)a^\dagger(0)\rangle^{(0)}$ are directed, and at zero-temperature the propagator vanishes for negative $\tau$. This greatly simplifies the perturbation theory since at zero-temperature all non-vanishing diagrams must involve arrows pointing in the same direction.
	
	The diagrams correcting the free energy at $\order{c^2}$ are shown in Fig \ref{Fig LL}. The strict requirements on the propagator at zero-temperature means that the bubble and melon diagrams correspond to far fewer scattering amplitudes than in Fig \ref{Fig bubble melon}.
	
	The free energy density due to the bubble diagram may be calculated straightforwardly in thermal field theory,
	\begin{align}
		f_{\text{bubble}}&=-2\beta c^2\left(\int \frac{dp}{2\pi }\frac{n(p)}{m}\right)^2\int \frac{dq}{2\pi}\frac{e^{-\beta E_q}}{\left(1-e^{-\beta E_q}\right)^2}.
	\end{align}
	From the DMB perspective this involves only the on-shell divergent three-body scattering amplitude depicted on the lower left in Fig \ref{Fig LL}. The result is a straightforward non-relativistic limit of the relativistic result \eqref{f bubble 3-body}.
	
	Note that the bubble diagram in LL does not have a two-body part. Recall that in the relativistic theory the two-body part of the bubble was associated with expanding the first-order in $c$ correction to the free energy density to the order $g^4$, but in the present case there is no $g^2$ parameter at all.
	
	As in the relativistic case, the bubble diagram may be found from the three-body part of the TBA \eqref{TBA free energy three body} if the principal value signs in the first order expansion of the kernel are ignored. The free energy density calculated from the TBA is
	\begin{align}
		f_{3}^{(2)}=-2\beta\frac{c^2}{m^2}\int\frac{dp_1dp_2dp_3}{\left(2\pi\right)^3}e^{\beta E_1}n_1^2n_2n_3\left[p_2 p_3\mathcal{P}\frac{1}{p_{12}}\mathcal{P}\frac{1}{p_{13}}+\dots_3\right].\label{TBA LL}
	\end{align}
	In this case quantities such as $p_{12}$ should be understood as $p_1-p_2$, and once again the identity \eqref{magic TBA identity} holds.
	
	The melon diagram involves a single two-body and a single three-body scattering amplitude depicted on the lower right of Fig \ref{Fig LL}. These can either be calculated directly, using the method of Appendix \ref{Sec Appendix melon}, or indirectly from the perturbative expansion of the TBA using \eqref{TBA free energy two body 2nd} and \eqref{TBA LL}. The result is a straightforward non-relativistic limit of the relativistic case \eqref{f melon 2-body} and \eqref{f melon 3-body},
	\begin{align}
		f_{\text{melon}}	&=-\frac{c^2}{12m}\int \frac{dp}{2\pi}\left(3n(p)^2+2n(p)^3\right).\label{f melon LL}
	\end{align}

%

\end{document}